\documentclass[preprint,aps,12pt,float,showpacs]{revtex4}
\usepackage{graphicx,color}
\usepackage{epsfig}
\usepackage{float}
\begin{document}
\begin{flushright}
SHEP-09-03\\
\end{flushright}
\newcommand  {\ba} {\begin{eqnarray}}
\newcommand  {\ea} {\end{eqnarray}}
\def\cM{{\cal M}}
\def\cO{{\cal O}}
\def\cK{{\cal K}}
\def\cS{{\cal S}}
\newcommand{\mh}{m_{h^0}}
\newcommand{\mw}{m_W}
\newcommand{\mz}{m_Z}
\newcommand{\mt}{m_t}
\newcommand{\mb}{m_b}
\newcommand{\be}{\beta}
\newcommand{\al}{\alpha}
\newcommand{\lam}{\lambda}
\newcommand{\no}{\nonumber}
\def\ga{\mathrel{\raise.3ex\hbox{$>$\kern-.75em\lower1ex\hbox{$\sim$}}}}
\def\la{\mathrel{\raise.3ex\hbox{$<$\kern-.75em\lower1ex\hbox{$\sim$}}}}


\title{Neutral Higgs boson pair production in photon-photon
annihilation in the Two Higgs Doublet Model}

\author{Abdesslam Arhrib$^{1,2}$, Rachid Benbrik$^{2,3,4}$,
Chuan-Hung  Chen$^{3,4}$  and Rui Santos$^{5}$}

\affiliation{$^1$ D\'epartement de Math\'ematique, Facult\'e des
Sciences and Techniques, Universit\'e Abdelmalek Essa\^adi, B.
416, Tangier, Morocco. } \affiliation{$^2$LPHEA, Facult\'e des
Sciences-Semlalia,
             B.P. 2390 Marrakesh, Morocco.}
\affiliation{$^3$ Department of Physics, National Cheng Kung
  University, Taiwan 701, Taiwan.}
\affiliation{$^4$ National Center for Theoretical Physics, Taiwan
701, Taiwan.} \affiliation{$^5$ NExT Institute and School of
Physics and Astronomy, University of Southampton Highfield,
Southampton SO17 1BJ, UK.}

\date{\today}

\begin{abstract}
We study double Higgs production in photon-photon collisions as a
probe of the new dynamics of Higgs interactions in the framework of
two Higgs Doublet Models. We analyze neutral Higgs bosons production
and decay in the fusion processes, $\gamma \gamma \to S_iS_j$,
$S_i=h^0, H^0, A^0$, and show that both $h^0h^0$ and $A^0A^0$
production can be enhanced by threshold effects in the region
$E_{\gamma\gamma}\approx 2 m_{H\pm}$.
Resonant effects due to the heavy Higgs, $H^0$, can also play a role in the cross section enhancement when it is allowed
to decay to two light CP-even $h^0$ or to two light CP-odd $A^0$
scalars. We have scanned the allowed parameter space of the Two
Higgs Doublet Model and found a vast region of the parameter space
where the cross section is two orders of magnitude above the Standard
Model cross section. We further show that the Standard Model experimental
analysis can be used to discover or to constraint the two Higgs doublet
model parameter space.

%
\end{abstract}


\maketitle

\section{Introduction}
The search for Higgs bosons is the prime task of CERN's Large
Hadron Collider (LHC), with operation scheduled now for 2009. With
the LHC guidance, the International $e^+ e^-$ Linear Collider
(ILC), which is currently being designed, will further improve our
knowledge of the Higgs sector if that is how Nature decided to
create mass. It was demonstrated in Ref.~\cite{ILCLHC} that
physics at the LHC and at the ILC will be complementary to each
other in many respects. In many cases, the ILC can significantly
improve the LHC measurements. If a Higgs boson is discovered, it
will be crucial to determine its couplings with high accuracy, to
understand the so-called mechanism of electroweak symmetry
breaking~\cite{Higgs:1964pj}. The high resolution profile
determination of a light Higgs boson (mass, couplings, self
couplings, etc.) can be carried out at the ILC, where clear
signals of Higgs events are expected with backgrounds that can be
reduced to a manageable level. This is exactly the case of
processes such as $e^+ e^- \to \gamma\gamma \to S_i S_j$ where
$S_i = h^0, H^0, A^0$. This fusion process can produce a Standard
Model (SM) Higgs boson or one predicted by the various extensions
of the SM, such as the Minimal Supersymmetric Standard Model
(MSSM) or Two Higgs Doublet Models (2HDM).

According to its Reference Design Report~\cite{ILC:2007sg}, the
ILC will run at an energy of $ \sqrt{s}= 500 $ $GeV$ with a total
luminosity of ${\mathcal{L}} = 500 fb^{-1}$ within the first four
years of operation and ${\mathcal{L}} = 1000 fb^{-1}$ during the
first phase of operation with $ \sqrt{s}= 500 $ $GeV$. An $e^+
e^-$ collider is uniquely capable of operation at a series of
energies near the threshold of a new physics process. This is an
extremely powerful tool for precision measurements of particle
masses and unambiguous particle spin determination. Various ILC
physics studies, indicate that a  $ \sqrt{s}= 500 $ $GeV$ collider
can have a great impact on understanding new physics at the $TeV$
scale. An energy upgrade up to $\sqrt{s} \sim 1$ $TeV$ would probably open
the doors to even greater discoveries. Another very unique feature
of the ILC is that it can accommodate a $\gamma\gamma$
collider with the photon beams generated by using the Compton
backscattering of the initial electron and laser
beams~\cite{Telnov:2006rj}. In this case, the energy and
luminosity of the photon beams would be of the same order of
magnitude of the original electron beams. As the set of final
states at a photon collider is much richer than that in the $e^+
e^-$ mode, it would open a wider window to probe new physics
beyond the SM.

Since photons couple directly to all fundamental fields carrying
electromagnetic charge, $\gamma \gamma $ collisions provide a
comprehensive means of exploring virtual aspect of the SM and its
extensions \cite{Boos:2001}. The production mechanism in hadron
and $e^+ e^-$ machines are often more complex and model-dependent.
Thus, a $\gamma \gamma $ collider is much more sensitive to new
physics even at higher mass scales \cite{gunion:1990}.

The primary mechanism of neutral Higgs boson production in
$\gamma\gamma$ collisions is $\gamma \gamma \to (h^0, H^0, A^0)$
\cite{Gunion,Borden,spira,Asner}, but in order to explore the
triple and quartic Higgs couplings at future high energy
colliders, it is necessary to study the Higgs boson pair
production process. The triple Higgs couplings of the 2HDM have
been extensively studied at $e^+e^-$ linear
colliders~\cite{linear} and shown to provide an opportunity to
measure those couplings. At photon-photon colliders, the cross
section for neutral Higgs boson pair production has been
calculated in \cite{jikia:1987,Belusevic:2004pz} in the SM and
found to be rather small. In the 2HDM, the process $\gamma \gamma
\to h^0h^0$ has been computed in the decoupling limit in
\cite{hollik:2008,knemaura}. They found that the cross section can
be substantially enhanced in the 2HDM and that the number of
events expected at the Photon Collider will allow a determination
or exclusion of some of the parameter space in the 2HDM potential.

In the MSSM, various studies for Higgs pair production at a photon
collider have been performed. The process $\gamma \gamma \to h^0
h^0$ was studied in \cite{Zhu:1997nz} while reactions $\gamma
\gamma \to h^0H^0, h^0A^0, H^0H^0, H^0A^0$ were determined in
\cite{Zhou:2003ss}. Process $\gamma \gamma \to A^0A^0$ was
calculated for the MSSM~\cite{Zhu:1998rh,Gounaris:2000ja} and
shown to have a cross section of the order of $0.1-0.2 \, fb$ for
a vast range of the photon-photon center of mass energy.

In this paper, we present a complete calculation of pair
production of all neutral Higgs bosons at the one loop level in
the 2HDM. We study the Higgs self couplings effects on the
$\gamma\gamma \to h^0 h^0$ and $\gamma\gamma \to A^0 A^0$
cross sections and briefly comment on
the $\gamma\gamma\to h^0A^0$, $\gamma \gamma \to h^0 H^0$,
$\gamma \gamma \to H^0A^0$ and
$\gamma \gamma \to H^0 H^0$ production modes. This exhausts all
possible neutral scalar production processes in the 2HDM. A
measurement of these processes can shed some light on the 2HDM
triple Higgs couplings. However, even if the situation regarding
a measurement of the vertex is not clear because no peak is detected,
a vast region of the 2HDM parameter space will be excluded.
The scalars will be detected via similar
final states because both the $h^0$ and the $A^0$, when not too
heavy, decay predominantly into fermions. In this regard, the
knowledge of their exact total cross section and angular
distributions may be helpful in order to distinguish between
CP-even and CP-odd scalars. Moreover, it is well-known that in the
2HDM, both the CP-even $h^0$ and the CP-odd pseudo-scalar $A^0$
can be rather light \cite{Opalpaper}. In fact, the bounds on the
$h^0$ and $A^0$ masses originate from the $e^+ e^- \rightarrow h^0
Z$ and $e^+ e^- \rightarrow h^0 A^0$ production processes with the
Higgs decaying to some combination of jets (mainly $b$ jets) and
$\tau$ leptons. The production process $e^+ e^- \rightarrow h^0 Z$
is proportional to $\sin^2 (\alpha-\beta)$ and this is the reason
why LEP does not limit the mass of a light Higgs $h^0$ for $\sin
(\alpha-\beta) = 0.1$. For $\sin (\alpha-\beta) = 0.3$ the bound
is of the order of 80 $GeV$ \cite{pdg4}. The pseudo-scalar mass is only
limited by the results on $e^+ e^- \rightarrow h^0 A^0$. However,
if the sum of the masses is above the LEP energy limit, again no
bound applies.

A very interesting feature of $\gamma \gamma \to A^0A^0$ is that a
light $A^0$ can easily emerge in the Next-to Minimal
Supersymmetric Standard Model (NMSSM) and therefore comparison
between models will certainly prove useful. In addition, we also
take into account in our calculation the perturbativity, unitarity
as well as vacuum stability constraints on the various parameters in the
Higgs potential.  We will show that after imposing those
constraints, cross sections are still large enough, in the
hundred of fempto-barn ($fb$) region in some cases, to probe the
2HDM scalar sector.  We will also study some of these
processes in the decoupling limit and in the fermiophobic limit of
the so-called type-I 2HDM.

The paper is organized as follows.  In the next section, we review
the 2HDM potential we will be using, give the analytical expressions
for the triple
and quartic Higgs couplings and list the theoretical constraints on the 2HDM
scalar potential such as unitarity and vacuum stability.
In Section~\ref{sec:cross sections},
we evaluate the double Higgs production cross section, $\gamma
\gamma \to S_i S_j$ with $S_{i,j}=H^0, \, h^0, \, A^0$, in the
general 2HDM paying special attention to $\gamma\gamma\to h^0 h^0$
and $\gamma\gamma\to A^0 A^0$. We then proceed to
Section~\ref{sec:numerical} where we present our numerical results
for the general 2HDM and for two limiting cases: the decoupling
limit and the fermiophobic limit. In Section~\ref{sec:signatures}
we discuss the final states in the different 2HDM scenarios. Our
findings are summarized in Section~\ref{sec:summary}.

\section{Review of the Two Higgs doublet model}
\subsection{The Two Higgs doublet model}
Two Higgs doublet models are some of the most well studied
extensions of the Standard Model. Various motivations for adding a
second Higgs doublet to the Standard Model have been advocated in
the literature \cite{Gun,abdel2}. There are several types of 2HDM.
While the coupling to gauge bosons is universal, there are many
ways to couple the Higgs doublets to matter fields. Assuming
natural flavor conservation \cite{Glashow:1976nt} there are four
ways to couple the Higgs to the fermions~\cite{berger}. The most
popular models are the type-I and the type-II models, denoted by
2HDM-I and 2HDM-II, respectively. In 2HDM-I, the quarks and
leptons couple only to one of the two Higgs doublet which is
exactly what happens in the SM.  In 2HDM-II, one of the 2HDM
fields couples only to down-type fermions (down-type quarks and
charged leptons) and the other one only couples to up-type
fermions in order to avoid the problem of flavor-changing neutral currents (FCNC's) at tree-level. There are two additional models less discussed in the literature: model III in which one of the doublets couples to all quarks and the other couples to all leptons and the type IV model is instead built such that one doublet couples to up-type quarks and to leptons and the other couples to down-type quarks. There is a class of models sometimes called also type-III and denoted as 2HDM-III where FCNC are induced at tree-level~\cite{rodrigues} which can lead to fine-tuning issues.
The discussion of the different models is not crucial as the
production processes discussed here have a very mild dependence on
the diagrams with fermion loops. It can however become relevant
when discussing the different final states. When the Higgs decays
predominantly to fermions, the relative size of the $h^0
\rightarrow b \bar{b}$, $h^0 \rightarrow c \bar{c}$ and $h^0
\rightarrow \tau^+ \tau^-$ branching ratios do depend on the model
chosen regarding the Yukawa sector. Finally, we note that a 2HDM-I
can lead to a fermiophobic Higgs boson $h^0$ \cite{landsberg_cite}
with suppressed couplings to the fermions (exactly zero at
tree-level). In this case, the dominant decay mode for the
lightest Higgs boson is $h^0\to \gamma \gamma$ or $h^0\to W^+W^-$,
depending on its mass. No other version of the 2HDM possesses such
a feature.

The most general scalar potential, renormalizable, CP-conserving,
invariant under $SU(2)_L\otimes U(1)_Y$ can be written
as~\cite{Gun}:
\ba V(\Phi_1,\Phi_2) &=& m^2_1 \Phi^{\dagger}_1\Phi_1+m^2_2
\Phi^{\dagger}_2\Phi_2 + (m^2_{12} \Phi^{\dagger}_1\Phi_2+{\rm
h.c}) +\frac{1}{2} \lam_1 (\Phi^{\dagger}_1\Phi_1)^2 +\frac{1}{2}
\lam_2 (\Phi^{\dagger}_2\Phi_2)^2\nonumber \\ &+& \lam_3
(\Phi^{\dagger}_1\Phi_1)(\Phi^{\dagger}_2\Phi_2) + \lam_4
(\Phi^{\dagger}_1\Phi_2)(\Phi^{\dagger}_2\Phi_1) + \frac{1}{2}
\lam_5[(\Phi^{\dagger}_1\Phi_2)^2+{\rm h.c.}] ~, \label{higgspot}
\ea
where $\Phi_1$ and $\Phi_2$ have weak hypercharge $Y=1$ and vacuum
expectation values (VEV's) $v_1$ and $v_2$, respectively, and
$\lambda_i$ and $m_{12}^2$ are real-valued parameters.  Note that
this potential violates the $Z_2$ discrete symmetry
$\Phi_1 \to \Phi_1$, $\Phi_2 \to - \Phi_2$
softly by the dimension-two term $m_{12}^2(\Phi^{\dagger}_{1}\Phi_{2})$,
and has the same general structure as the scalar potential in the MSSM.
From hermeticity of the potential, one concludes that $m_{12}^2$ must be
real valued, so it can take both positive and negative values.

After electroweak symmetry breaking, the $W^\pm$ and $Z$ gauge
bosons acquire their masses.  Explicitly, three of the eight
degrees of freedom in the two Higgs doublets correspond to the
three Goldstone bosons ($G^\pm$, $G^0$) and the remaining five
become physical Higgs bosons: $h^0$, $H^0$ (CP-even), $A^0$
(CP-odd), and $H^\pm$ with masses $m_{h^0}$, $m_{H^0}$, $m_{A^0}$,
and $m_{H^\pm}$, respectively.

The potential in Eq.~(\ref{higgspot}) has ten independent
parameters (including $v_1$ and $v_2$).  The parameters $m_1$ and
$m_2$ are fixed by the minimization conditions.  The combination
$v^2=v_1^2 + v_2^2$ is fixed as usual by the electroweak breaking
scale through $v^2=(2\sqrt{2} G_F)^{-1}$.  We are thus left with
seven independent parameters; namely $(\lambda_i)_{i=1,\ldots,5}$,
$m_{12}$, and $\tan\beta \equiv v_2/v_1$.  Equivalently, we can
take instead 
\ba m_{h^0}\quad , \quad m_{H^0} \quad , \quad m_{A^0} \quad ,
\quad m_{H^\pm} \quad , \quad \tan\beta \quad , \quad \alpha \quad
\rm{and} \quad m_{12} \, , \label{parameters} \ea 
as the seven independent parameters.  The angle $\beta$
diagonalizes both the CP-odd and charged scalar mass matrices and
$\alpha$ diagonalizes the CP-even mass matrix.  One can easily
relate the physical scalar masses and mixing angles from
Eq.~(\ref{higgspot}) to the potential parameters, $\lambda_i$,
$m_{12}$ and $v_i$, and invert them to obtain $\lambda_i$ in terms
of the physical scalar masses, $\tan\beta$, $\alpha$, and $m_{12}$
\cite{Gunion:2002zf,am}.


\subsection{Theoretical and experimental constraints}

There are several important constraints on the 2HDM parameters
imposed by experimental data. In our analysis we take them all into
account when the independent parameters are varied.

First, the LEP direct search result in the lower bounds $m_{h^0}
>$ 114 $GeV$ for a SM-like Higgs and $m_{A^0, H^0, H^\pm} > $ 80-90
$GeV$  for supersymmetric models in the case of the neutral
scalars and for more general models in the case of the charged
Higgs (see \cite{pdg4} for details). As stated in the
introduction, the bound on the lightest CP-even Higgs heavily
depends on the value of $\sin (\alpha- \beta)$. In a general 2HDM
all bounds on the Higgs masses, with the exception of the charged
Higgs, can be avoided with a suitable choice of the angles and
$m_{12}$.

Second, the extra contributions to the $\delta\rho$ parameter from
the Higgs scalars \cite{Rhoparam} should not exceed the current
limit from precision measurements \cite{pdg4}: $ |\delta\rho| \la
10^{-3}$. Such an extra contribution to $\delta\rho$ vanishes in
the limit $m_{H^\pm}=m_{A^0}$.  To ensure that $\delta\rho$ is
within the allowed range, we demand either a small splitting
between $m_{H^\pm}$ and $m_{A^0}$ or a combination of parameters
that produces the same effect.

Third, the constraint from $B\to X_s \gamma$ branching ratio
\cite{Oslandk,Misiak} gives a lower bound  on the charged Higgs
mass, $m_{H^\pm} \ga 295$ $GeV$, in 2HDM-II. These bounds do not
apply to model type-I and therefore are not taken into account in
the fermiophobic scenario. Recent data from $B \to \ell \nu$ can
also give a constraint on charged Higgs mass especially for large
values of $\tan\beta$ in 2HDM-II \cite{chen1,ikado}.

Fourth, values of $\tan \beta$ smaller than $\approx 1$ are
disallowed both by the constraints coming from $Z \rightarrow b
\bar{b}$ and from $B_q \bar{B_q}$ mixing~\cite{Oslandk}.

Finally, we should take into account the theoretical constraints.
Let us start by noting that all 2HDM are protected against charge and
CP-breaking~\cite{charge}. We consider the perturbativity constraints
on the $\lambda_i$ as well as the vacuum stability conditions
\cite{vac1} that assure that the potential is bounded from below.
We require that all quartic couplings of the scalar potential
Eq.~\ref{higgspot} remain perturbative by imposing
$|\lambda_i| \leq 8 \pi$ for all $i$. For the vacuum stability
conditions we use those from \cite{vac1}, which are given by:
\begin{eqnarray}
\nonumber
& \lambda_1  > 0\;,\quad\quad \lambda_2 > 0\;,
\nonumber\\
& \sqrt{\lambda_1\lambda_2 }
+ \lambda_{3}  + {\rm{min}}
\left( 0 , \lambda_{4}-|\lambda_{5}|
 \right) >0  \, \, . \label{vac}
\end{eqnarray}
The above perturbative constraints are slightly less constraining
than the full set of unitarity constraints \cite{unit1,abdesunit}
established using the high energy approximation as well as the
equivalence theorem. It turns out, that requiring only perturbativity
constraints on the $\lambda's$ could lead to scalar particles having a
decay width which could exceed their mass. The problem is cured when we use the full set of perturbative unitarity conditions
which are given by
\begin{eqnarray}
  |a_{\pm}|,  |b_{\pm}|,  |c_{\pm}|,  |d_{\pm}|,
   |e_{1,2}^{}|,  |f_{\pm}|,  |g_{1,2}^{}|  < 8 \pi \label{unita}
\end{eqnarray}
with
\begin{eqnarray}
a_{\pm}^{} &=&
 \frac{3}{2} \left\{
  (\lambda_1 + \lambda_2) \pm
   \sqrt{ (\lambda_1-\lambda_2)^2 + \frac{4}{9} (2\lambda_3+\lambda_4)^2}
  \right\},  \\
b_{\pm}^{} &=&
 \frac{1}{2} \left\{
   (\lambda_1 + \lambda_2) \pm
   \sqrt{ (\lambda_1-\lambda_2)^2 +4 \lambda_4^2}
  \right\},  \\
c_{\pm}^{} &=& d_{\pm}^{}=
 \frac{1}{2} \left\{
   (\lambda_1 + \lambda_2) \pm
   \sqrt{(\lambda_1-\lambda_2)^2 +4 \lambda_5^2}
  \right\},  \\
e_1 &=&    \left(
    \lambda_3 + 2 \lambda_4 - 3 \lambda_5 \right) \qquad , \qquad
e_2 =    \left(
    \lambda_3 - \lambda_5 \right), \\
f_+ &=&    \left(
    \lambda_3 + 2 \lambda_4 + 3 \lambda_5 \right) \qquad , \qquad
f_- =    \left(
    \lambda_3 + \lambda_5 \right), \\
g_1 &=& g_2 =    \left(
    \lambda_3 + \lambda_4 \right).
\end{eqnarray}
These are very restrictive constraints on the allowed
range of the parameter space. All values
presented in the plots are consistent with all theoretical and
experimental bounds described in this section.
%
\section{$\gamma\gamma \to S_iS_j$,
$S_{i,j}=h^0,H^0,A^0$ in the  2HDM }
\label{sec:cross sections}

\subsection{About the one-loop calculation}

\begin{figure}[h!]
\begin{center}
\epsfig{file=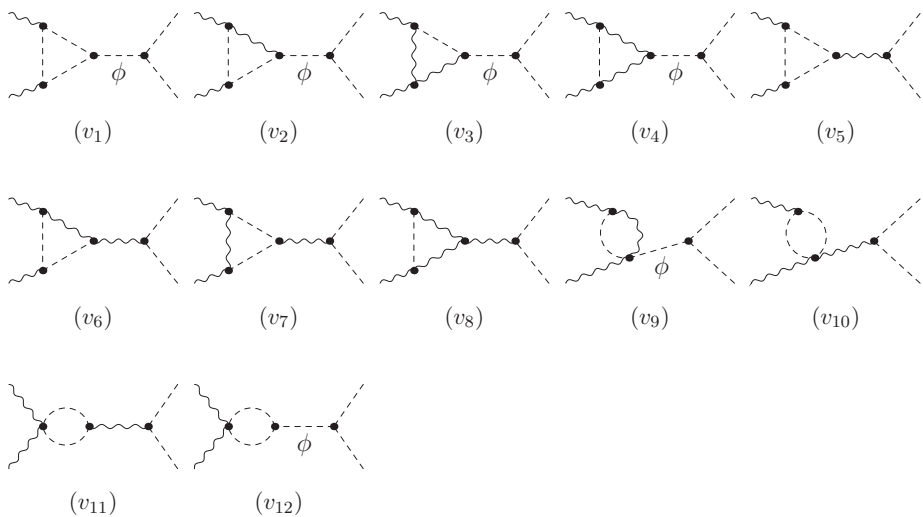,width=14 cm}
\caption{Generic charged Higgs and gauge bosons
vertex like Feynman diagrams for neutral Higgs production $\gamma
\gamma \to S_iS_j$ in 2HDM.  In the figures $\phi=h^0$ or $H^0$.}
\label{fig:vert-diagrams}
\end{center}
\end{figure}

\begin{figure}[h!]
  \begin{center}
    \epsfig{file=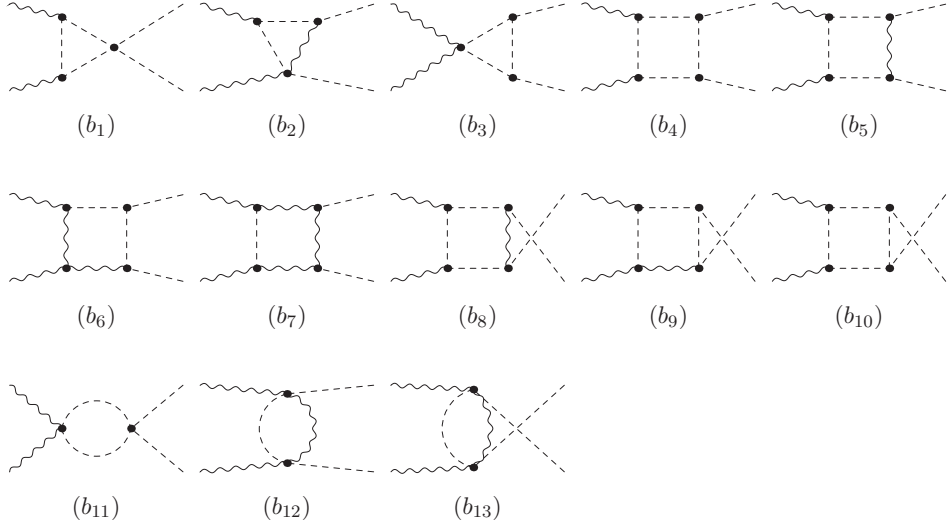,width=14 cm}
    \caption{Generic charged Higgs and gauge bosons box
like Feynman diagrams to neutral Higgs production $\gamma \gamma \to
S_iS_j$ in 2HDM.} \label{fig:box-diagrams}
  \end{center}
\end{figure}

All processes $\gamma\gamma \to S_iS_j$, $S_{i,j}=h^0,H^0, A^0$
with neutral Higgs in the final state are forbidden at tree-level
and are mediated at one-loop level by vertex diagrams as well as
by box diagrams. All those processes are sensitive to virtual
gauge bosons, fermions and charged Higgs particles. We display in
Fig.~\ref{fig:vert-diagrams} and in Fig.~\ref{fig:box-diagrams}
the generic Feynman diagrams with charged scalar particles
exchange that contribute to  $\gamma \gamma \to S_iS_j$ processes.
Note that in Fig.~\ref{fig:vert-diagrams} and in
Fig.~\ref{fig:box-diagrams} we do not show the SM contribution
with fermions and gauge bosons exchange. We have checked that they
are always negligible when compared to diagrams with scalar
exchange. The later comprise one-loop photon-photon fusion
diagrams resulting in $\phi=h^0$ or $H^0$ intermediate states,
followed by the decay $h^0,H^0 \to S_iS_j$ with $S_{i,j}=h^0,A^0$
as shown in Fig.~\ref{fig:vert-diagrams}, $v_{1\rightarrow 4}$,
$v_{9}$ and $v_{12}$. This kind of topology is sensitive to the
triple Higgs couplings $h^0S_iS_j$, $H^0S_iS_j$, $h^0H^+H^-$ and
$H^0H^+H^-$. For the $\gamma \gamma \to h^0A^0$ process, we have
diagrams like $v_{1\rightarrow 4}$, $v_{9}$ and $v_{12}$ but with
$\phi=A^0$ and also
 the contribution from the vertices with s-channel exchange of a $Z$
boson $v_{5 \rightarrow 8}$, $v_{10}$ and $v_{11}$. Note that for
topologies like $v_{1\rightarrow 4}$, $v_{9}$ and $v_{12}$ in
Fig.~\ref{fig:vert-diagrams}  we have included the total width of
the scalar particle $\phi$ in the calculation of the corresponding
amplitude.

For the production mode $\gamma\gamma\to h^0h^0$, the box
contributions with virtual charged Higgs exchange is sensitive to
the triple Higgs coupling $h^0H^+H^-$. On the contrary, in the
case of the $\gamma \gamma \to A^0A^0$ process and again due to
the CP nature of the pseudo-scalar Higgs boson $A^0$, it turns out
that the box diagrams for $\gamma\gamma\to A^0A^0$ are rather
sensitive to the $A^0H^+G^-$ coupling which does not have neither
a $m_{12}$ nor a $\tan\beta$ dependence. As one can see from
Fig.~\ref{fig:box-diagrams} (diagrams ($b_1$) and ($b_{11}$)),
there are other topologies that contribute to $\gamma\gamma\to
h^0h^0$ and $\gamma\gamma\to A^0A^0$ and which are sensitive to
quartic couplings of the Higgs boson such as $h^0h^0H^+H^-$ and
$A^0A^0H^+H^-$.

As stated before, we are mainly concerned with the production
modes $\gamma \gamma \to h^0h^0$ and $\gamma \gamma \to A^0 A^0$.
We will briefly comment on the $h^0A^0$, $H^0A^0$, $h^0 H^0$ and
$H^0 H^0$ production processes. The one-loop amplitudes were
generated and calculated with the packages FeynArts \cite{FA} and
FormCalc \cite{FC}. The scalar integrals were evaluated with
LoopTools \cite{LT}. The numerical evaluations of the integration
over $2 \to 2$ phase space is done by the help of CUBA library
\cite{cuba}. A cut of approximately $6^o$ relative to the beam
axis was set on the scattering angle in the forward and backward
directions.

\subsection{Triple Higgs couplings}
The above processes are sensitive to triple and quartic Higgs
couplings. Below, we list the relevant pure scalar couplings
needed for our processes $\gamma\gamma\to h^0h^0, A^0A^0, h^0A^0$.
In the SM and in the general 2HDM these triple and quartic scalar
couplings are given at tree-level by
\begin{eqnarray}
\label{trip1}
\lambda_{h^0h^0h^0}^{SM} & = & \frac{-3 g m_{h^0}^2}{2 m_W } \quad , \quad
\lambda_{h^0h^0h^0h^0}^{SM}  =  \frac{-3 g^2 m_{h^0}^2}{4 m_W^2 }
\label{hhhsm}\\
\lambda_{h^0h^0h^0}^{2HDM} &=&
\frac{-3 g}{m_W s^2_{2\be}}\bigg[(c_\be c^3_\al - s_\be s^3_\al)s_{2\be} m^2_{h^0} + c^2_{\be-\al} c_{\be + \al} m^2_{12}\bigg]
\label{lll}\\
\lambda_{H^0H^0H^0}^{2HDM} &=& \frac{-3 g }{m_W s^2_{2\be}}\bigg[(c_\be c^3_\al - s_\be
  s^3_\al)s_{2\be} m^2_{H^0} + s^2_{\be-\al} s_{\be + \al} m^2_{12}\bigg]
\label{HHH}  \\
\lambda_{H^0h^0h^0}^{2HDM} &=& -\frac{1}{2}\frac{g c_{\be-\al}}{s_W s^2_{2\be}}\bigg[
  (2 m^2_{h^0} + m^2_{H^0}) s_{2\al} s_{2\be} + (3 s_{2\al}-s_{2\be})
  m^2_{12}\bigg]\label{hll} \\
\lambda_{H^0H^0h^0}^{2HDM} &=& \frac{1}{2}\frac{g s_{\be-\al}}{m_W s^2_{2\be}}\bigg[
  (m^2_{h^0} + 2 m^2_{H^0}) s_{2\al} s_{2\be} + (3 s_{2\al}+s_{2\be})
  m^2_{12}\bigg] \label{hhl}\\
\lambda_{A^0A^0h^0}^{2HDM} &=& \frac{-g}{m_W s^2_{2\be}}\bigg[(c_\al c^3_\be - s_\al
  s^3_\be)s_{2\be} m^2_{h^0} + c_{\be+\al} m^2_{12}  + s^2_{2\be}
  s_{\be-\al} m^2_{A^0} \bigg]\label{aal}\\
\lambda_{A^0A^0H^0}^{2HDM} &=& \frac{-g}{m_W s^2_{2\be}}\bigg[(s_\al c^3_\be + c_\al
  s^3_\be)s_{2\be} m^2_{H^0} + s_{\be+\al}m^2_{12} +s^2_{2\be}
  c_{\be-\al} m^2_{A^0} \bigg]\label{aah}\\
\lambda_{A^0G^0h^0}^{2HDM} &=& \frac{1}{2}\frac{g c_{\be-\al}}{m_W}\bigg( m^2_{A^0}
- m^2_{h^0}\bigg)\quad , \quad \lambda_{A^0G^0H^0}^{2HDM} = -\frac{1}{2}\frac{g s_{\be-\al}}{m_W}\bigg( m^2_{A^0}
- m^2_{H^0}\bigg)\label{agh}\\
\lambda_{H^\pm H^\mp h^0}^{2HDM}&=& \frac{g}{m_W s_{2\be}}\bigg[(s_\al s^3_\be - c_\al
  c^3_\be)s_{2\be} m^2_{h^0} - c_{\be+\al} m^2_{12} - s^2_{2\be}
  s_{\be-\al} m^2_{H^\pm} \bigg]\label{lhphm}\\
\lambda_{H^\pm H^\mp H^0}^{2HDM} &=& \frac{-g}{m_W s_{2\be}}\bigg[(s_\al c^3_\be + s_\al
  s^3_\be)s_{2\be} m^2_{H^0} + s_{\be+\al} m^2_{12} + s^2_{2\be}
  c_{\be-\al} m^2_{H^\pm} \bigg]\label{hhphm}\\
 \lambda^{2HDM}_{h^0 h^0 H^- H^+} &=\nonumber & -
\bigg(\frac{g}{2 m_W s_{2\beta}}\bigg)^2\bigg[ 2 m^2_{H^\pm} s^2_{2\beta}
s^2_{\beta-\alpha} + m^2_{h^0} (2 c_{\alpha+\beta} +
s_{2\alpha}s_{\beta-\alpha})(2 c_{\alpha+\beta} -
s_{2\beta}s_{\beta-\alpha}) \\
&-&
c_{\beta-\alpha}s_{2\alpha}(c_{\beta - \alpha} s_{2\beta} -
2s_{\alpha+\beta}) m^2_{H^0} +  m^2_{12}
(c^2_{\alpha+\beta} + c^2_{2\beta} c^2_{\beta-\alpha})\bigg].\\
\lambda^{2HDM}_{A^0 A^0 H^- H^+} &=& -\bigg(\frac{g}{2
m_{W}s_{2\beta}}\bigg)^2
\bigg[ m^2_{H^0}
(c_{\beta-\alpha}s_{2\beta}-2s_{\beta+\alpha})^2 + m^2_{h^0}
(2c_{\beta+\alpha} - s_{2\beta}s_{\beta-\alpha})^2
\\\nonumber&+&2m^2_{12} c^2_{2\beta}\bigg]
 \end{eqnarray}
where $g= e/\sin\theta_W$ is the $SU(2)_L$ gauge coupling
constant. Here we use the short-hand notations $s_{\theta}$ and
$c_{\theta}$ to denote, respectively, $\sin\theta$ and
$\cos\theta$ where $\theta$ stands for $\alpha$ or $\beta$.
All these triple Higgs
couplings have a strong dependence on the physical masses
$m_{\phi}$ ($\phi = h^0, H^0, H^\pm, A^0$), on the mixing angles
$\alpha$ and $\beta$ and finally on the  $m_{12}$ parameter which
parameterizes the soft breaking of the $Z_2$ symmetry.

\begin{figure}[!ht]
\centering
\includegraphics[height=3.0in,width=7cm]{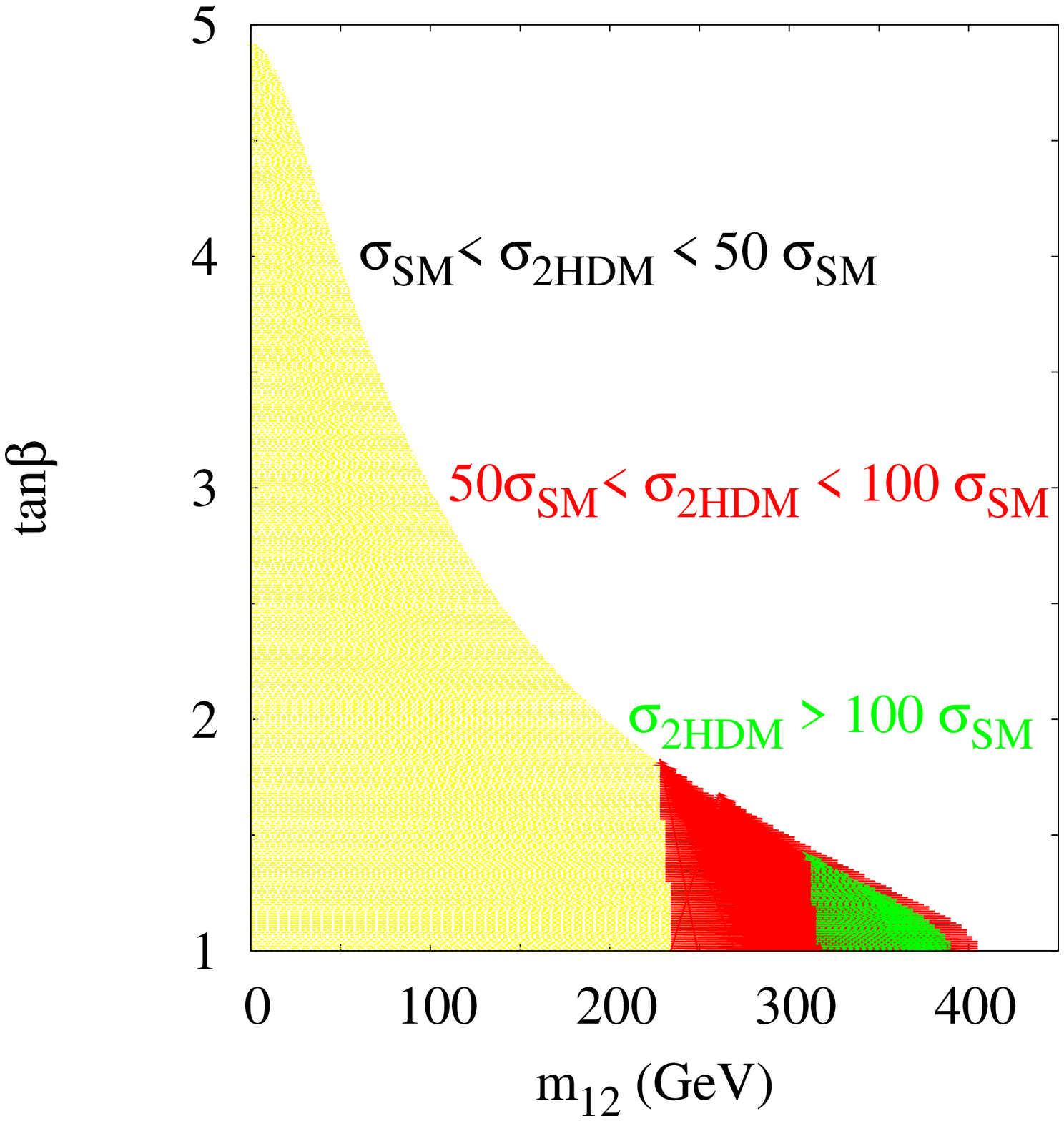}
\hskip 0.1cm
\includegraphics[height=3.0in,width=7cm]{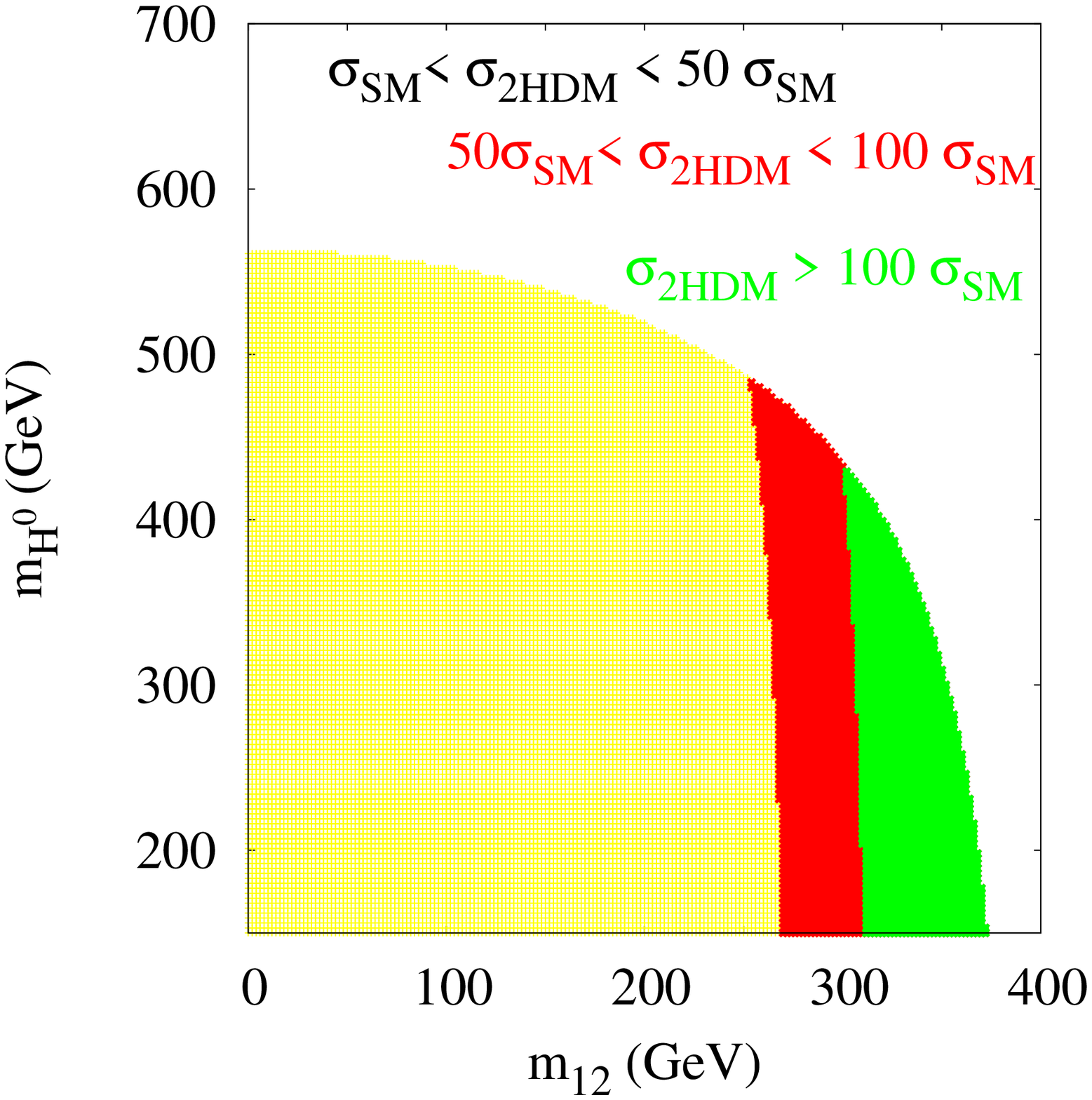}
\caption{The allowed regions for $\sigma(\gamma\gamma \to h^0 h^0)$
in two Higgs doublet model. We have chosen $m_{h^0}= 115$ GeV, $m_{A^0}= 270$ GeV 
and $m_{H^\pm} = 350$ GeV. On the left panel, $m_{H^0} = 2 m_{h^0}$, $E_{\gamma \gamma} = 500$ GeV, $-1 \le \sin\alpha \le 1$ and $1\la \tan\beta\la 10$. On the right panel, 
$\tan\beta =1, \, \sin\alpha = -0.9$ and $E_{\gamma \gamma} = 800$ GeV }
\label{scan1}
\end{figure}
\section{Numerical results}
\label{sec:numerical}

\subsection{The general 2HDM}
Before discussing  our numerical results, it is worth pointing out that
the following results are valid for all Yukawa type of couplings that do
not generate FCNC at tree-level, as long as
$\tan\beta$ remains small ($\tan\beta\la 7$ in the regions probed),
as imposed by unitarity constraints. Moreover, as we will see later,
the 2HDM contribution is dominated by scalar loops rather than by
fermion loops and the former are Yukawa model independent.
Since data can easily accommodate light $h^0$ and $A^0$
scalars~\cite{Opalpaper} in the 2HDM, we will concentrate
hereafter on the $h^0h^0$, $A^0A^0$ and $h^0A^0$ production modes. In our numerical analysis, we will use:
$m_t=171$ GeV, $m_b=4.7$ GeV, $m_Z=91.187$ GeV, $m_W=80.45$ GeV,
the Weinberg angle $s_W$ is defined in the on-shell scheme as
$s_W^2=1-m_W^2/m_Z^2$.
For the fine structure constant we use $\alpha=1/137.035989$.

We first note that we have reproduced the SM result for
$\gamma\gamma\to H^0H^0$ and found perfect agreement with
\cite{jikia:1987,Belusevic:2004pz}. In the 2HDM case, our result agree
with \cite{hollik:2008} while we have a full
agreement with \cite{knemaura} if we take $\alpha=1/128$.
The very detailed parton-level study~\cite{Belusevic:2004pz}
conclude that for a 350 $GeV$ center of
mass energy photon collider and a Higgs mass of 120 $GeV$, an integrated
$\gamma \gamma$ luminosity of 450 $fb^{-1}$ would be needed to exclude a
zero trilinear Higgs boson self-coupling at the $5\sigma$ level,
considering only the statistical uncertainty. If one assumes the
luminosity based on the TESLA design report~\cite{Badelek:2001xb}
we conclude that this is an attainable luminosity in approximately one year
and certainly in less than two years. Therefore, we have decided to
perform a comprehensive scan of the parameter space of the 2HDM looking
for regions where the 2HDM dominate over the SM, that is,
$\sigma_{2HDM}(\gamma\gamma \to h^0 h^0) >
\sigma_{SM} (\gamma\gamma \to h^0 h^0) $, together with perturbativity,
unitarity and vacuum stability constraints on $\lambda_i$.
The results of this scan are shown in  Fig.~\ref{scan1}.
In this scan we choose the charged Higgs mass to be 350 GeV,
in order to fulfill the $b\to s\gamma$ constraint in all Yukawa type models.
From the left scan, it comes out a correlation between
$\tan\beta$ and $m_{12}$. In order to have $m_{12}$ large,
unitarity constraints require $\tan\beta$ to be rather small.
It is clear that in order to have a 2HDM cross section for
$\gamma\gamma \to h^0h^0$ much larger than the corresponding SM one,
a large $m_{12}$ is needed together with a small value for
$\tan\beta\approx 1,1.5$. With the set of parameters fixed,
unitarity constraints forces $\tan\beta\la 5$ in the scanned region.
The left scan shows that as
$\tan\beta$ grows larger and larger, only smaller values of
$m_{12}$ are allowed.
In the right panel of  Fig.~\ref{scan1}
one can see that a quite large range of
$m_H$ is allowed by perturbativity, unitarity
and vacuum stability constrains.
For large $m_{12}\ga 260$ GeV and
for all values of $m_H\la 500$ GeV, the 2HDM cross section for
$\gamma\gamma \to h^0 h^0$
can be 50 times larger than the corresponding SM cross section.
From the scans presented, we conclude that for a large
$m_{12}$ and even if $\tan\beta \approx O (1)$ we
have a significant slice of the parameter space where
$\sigma_{2HDM}(\gamma\gamma \to h^0 h^0)$ can be much larger
than the corresponding SM cross section while complying with
all constraints both experimental and theoretical.
\begin{widetext}
\begin{figure}
\begin{center}
\includegraphics[width=7.5cm]{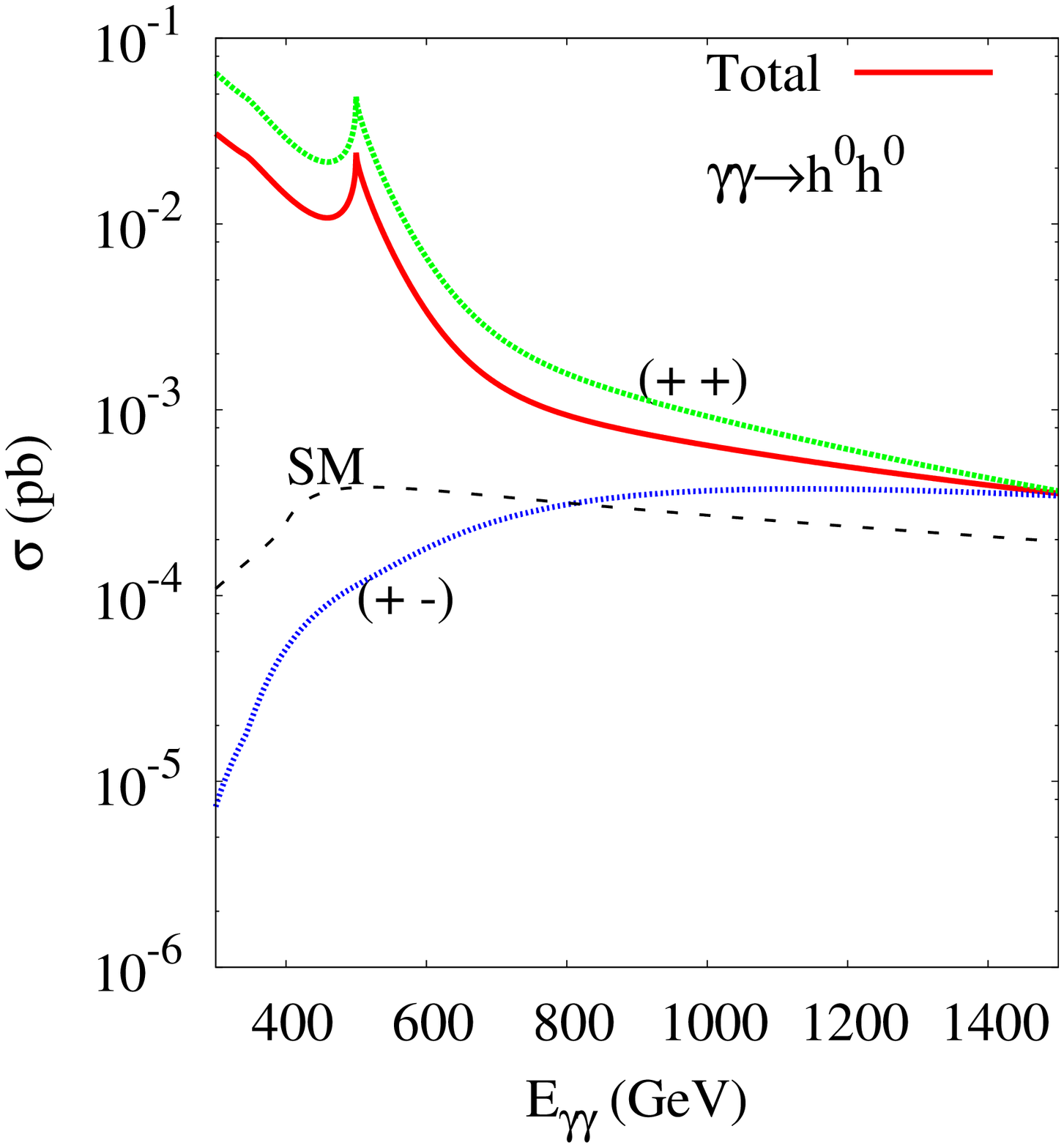}
\includegraphics[width=7.5cm]{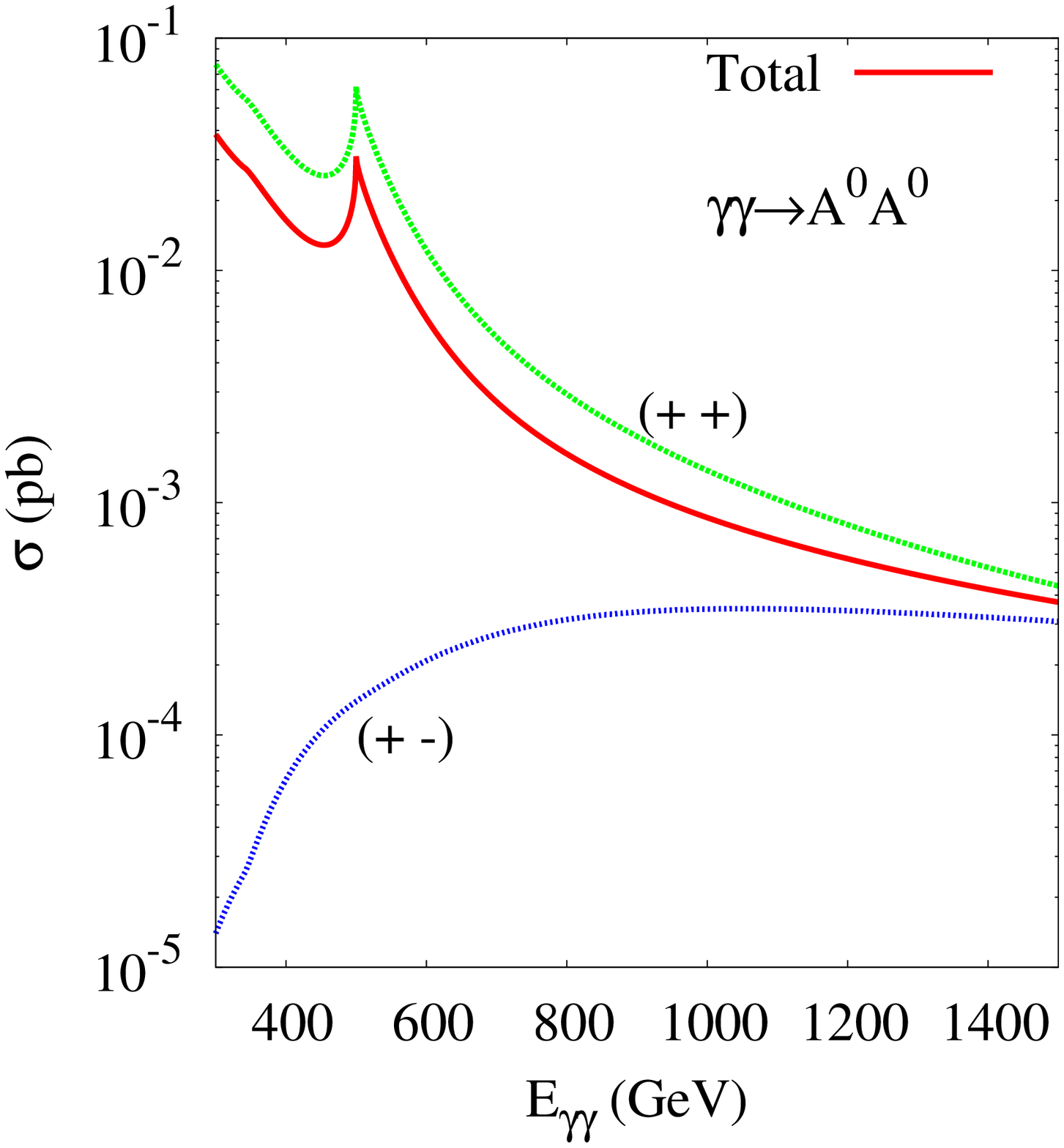}
\end{center}
\caption{SM and 2HDM total cross
sections $\sigma(\gamma\gamma \to h^0h^0)$ (left) and
$\sigma(\gamma\gamma \to A^0A^0)$ (right) as a function of the two
photon center of mass energy for unpolarized beams. Also shown is
the total cross section in two other situations: both beams with
right polarization and the two beams with opposite polarization.
The parameters are chosen to be
$m_{h^0}, m_{H^0}, m_{A^0}, m_{12} = 115, 130, 100, 400$ GeV,
$\sin\alpha = 0.6$, $\tan\beta = 1$ and  $m_{H^\pm} = 250$ GeV. }
\label{fig1}
\end{figure}
\end{widetext}

In Fig.~\ref{fig1} (left) we show the polarized and unpolarized cross section
for $\gamma\gamma\to h^0h^0$ both in the SM and in the 2HDM.
 The cross section is amplified by the threshold
effect when $E_{\gamma\gamma}\approx 2m_{H\pm}=500$ $GeV$,
corresponding to the opening of the charged Higgs pair channel,
$\gamma\gamma\to H^+H^-$. Near this threshold region, the cross
section of the 2HDM is more than two orders of magnitude larger
than the SM one.
Note that the s-channel vertex contribution  is suppressed for large values
of the center of mass energy due to the s-channel propagator. 
At high energies, box diagrams will dominates over the s-channel vertex.
%
%
Let us take $h^0 h^0$ production as an
example. The value of the couplings for this set of parameters is
\begin{eqnarray}
&&\lambda_{h^0H^+H^-}^{2HDM}\approx 3\times \lambda_{h^0h^0h^0}^{SM}
\quad , \quad \lambda_{h^0A^0A^0}^{2HDM}\approx
2.5\times \lambda_{h^0h^0h^0}^{SM}
\nonumber\\
&& \lambda_{H^0H^+H^-}^{2HDM}\approx 20\times \lambda_{h^0h^0h^0}^{SM}
\quad , \quad
\lambda_{H^0A^0A^0}^{2HDM}  \approx
17.5\times \lambda_{h^0h^0h^0}^{SM} \nonumber\\
&& \lambda_{h^0h^0h^0}^{2HDM}\approx
7 \times \lambda_{h^0h^0h^0}^{SM} \quad , \quad
\lambda_{H^0h^0h^0}^{2HDM}
\approx 16.5\times \lambda_{h^0h^0h^0}^{SM}\nonumber\\
&& \lambda_{h^0h^0H^+ H^-}^{2HDM}\approx \lambda_{h^0h^0h^0h^0}^{SM}
\quad , \quad
\lambda_{A^0A^0H^+ H^-}^{2HDM}\approx
0.5\times \lambda_{h^0h^0h^0h^0}^{SM} \, \, .
\end{eqnarray}
Now the largest contribution to the vertex comes from the
$H^0H^+H^-$  and $H^0h^0h^0$ couplings squared which are
approximately 400 and 272 times larger than
$(\lambda_{h^0h^0h^0}^{SM})^2$.
Note that in the case of $A^0A^0$ there are no boxes
with virtual $H^\pm$ exchange because the $A^0H^+H^-$ coupling does
not exist and consequently the vertex contribution will
dominate over the box contribution.

In the right panel of Fig.~\ref{fig1} the CP-odd Higgs boson pair
production $\gamma\gamma\to A^0A^0$ is shown.
The conclusions are very similar to the ones for the $h^0 h^0$
final state. The total cross sections is dominated
by vertex contributions for low energy.
This dominance is amplified by the threshold effect when
$E_{\gamma\gamma}\approx 2m_{H\pm}$.
For high center of mass energies, box contributions
dominate because they have those $t$ and $u$ channel topologies which are
enhanced for large center of mass energies. Finally, we show that when the initial photons are both right handed $(++)$ or left handed $(--)$ polarized, the cross sections are enhanced by more than a factor two as compared to the unpolarized case.
In the opposite case, when the initial
photons have opposite polarizations, $(+-)$ or $(-+)$, the cross sections are now suppressed when compared to unpolarized case. 

\begin{widetext}
\begin{figure}
\begin{center}
\includegraphics[width=7.5cm]{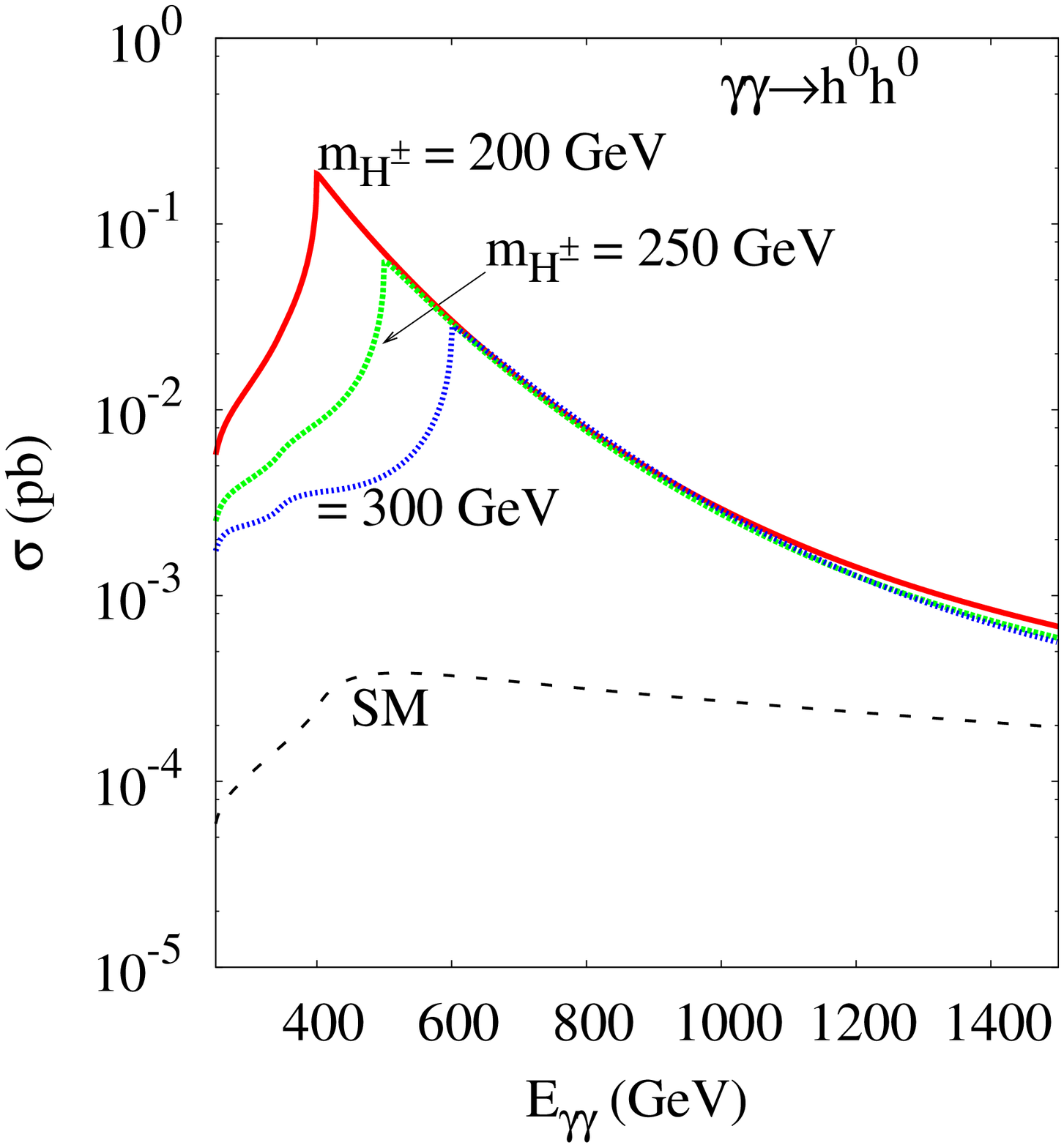}
\includegraphics[width=7.5cm]{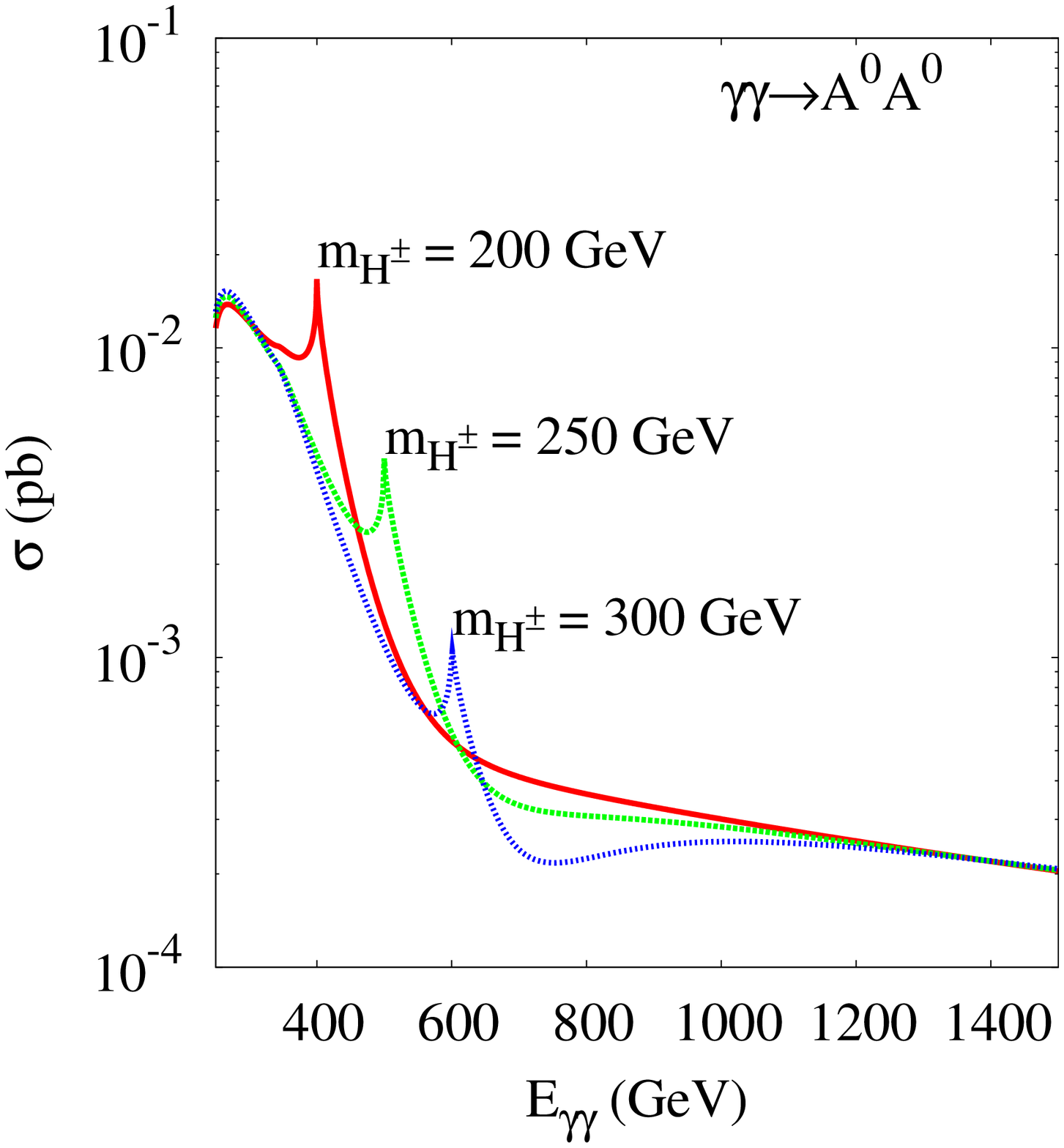}
\end{center}
\caption{The total cross section of $\sigma(\gamma\gamma \to hh)$
(left) and $\sigma(\gamma\gamma \to A^0A^0)$ (right) as a function of
two photon center of mass energy in the 2HDM. With $m_{h^0},
m_{H^0}, m_{A^0}, m_{12} = 120, 200, 120, 300$ GeV,
$\sin\alpha = -0.86$ and $\tan\beta = 1$ for different values of $m_{H^\pm}$. }
\label{fig2}
\end{figure}
\end{widetext}

The box contribution has
either the quartic vertex $h^0 h^0 H^+H^-$ or twice the triple
vertex $h^0H^+H^-$ which means that the cross section will have
$(\lambda_{h^0H^+H^-})^4$ dependence.
When $h^0H^+H^-$ and/or $h^0 h^0 H^+H^-$ are large
enough the boxes could dominate over the vertices.
By tuning $\sin\alpha$ one can make $h^0H^+H^-$ and $h^0 h^0 H^+H^-$
large enough to allow for the boxes to dominate. This is exactly what
is illustrated in the following Fig.~\ref{fig2},
where we present the total cross section for
$\gamma\gamma \to h^0h^0$ (left) and $\gamma\gamma \to A^0A^0$
(right) as a function of the two photon center of mass energy for different values of the charged Higgs mass. Once more we can see the enhancement when $E_{\gamma\gamma}\approx 2 m_{H\pm}$ for $\gamma\gamma \to h^0h^0$ and the cross section can reach 0.2 pbarn. For this specific scenario,
the triple and quartic couplings are
\begin{eqnarray}
&&\lambda_{h^0H^+H^-}^{2HDM}\approx 13\times \lambda_{h^0h^0h^0}^{SM}
\quad , \quad \lambda_{h^0A^0A^0}^{2HDM}\approx
9\times \lambda_{h^0h^0h^0}^{SM}
\nonumber\\
&& \lambda_{H^0H^+H^-}^{2HDM}\approx 4\times \lambda_{h^0h^0h^0}^{SM}
\quad , \quad
\lambda_{H^0A^0A^0}^{2HDM}  \approx
2\times \lambda_{h^0h^0h^0}^{SM} \nonumber\\
&& \lambda_{h^0h^0h^0}^{2HDM}\approx
3 \times \lambda_{h^0h^0h^0}^{SM} \quad , \quad
\lambda_{H^0h^0h^0}^{2HDM}
\approx 3\times \lambda_{h^0h^0h^0}^{SM}\nonumber\\
&& \lambda_{h^0h^0H^+ H^-}^{2HDM}\approx 12 \times \lambda_{h^0h^0h^0h^0}^{SM}
\quad , \quad
\lambda_{A^0A^0H^+ H^-}^{2HDM}\approx
0.5 \times \lambda_{h^0h^0h^0h^0}^{SM}
\end{eqnarray}
In the case of the $\gamma\gamma\to h^0h^0$
mode, the total cross section is now fully dominated by box
contributions both at low and high energies.
This is because the couplings $h^0H^+H^-$ and $h^0h^0H^+H^-$
which contribute to the boxes are relatively large.
In the case of $\gamma\gamma\to A^0A^0$ mode, for low energy,
the total cross section is dominated by vertex diagrams because the triple couplings $h^0H^+H^-$ and $h^0A^0A^0$
 are large. For high energies, where vertex are suppressed,
the total cross section is dominated by the box contributions.
Hence, at high energies, both for the
$\gamma\gamma\to h^0h^0$ and for the $\gamma\gamma\to A^0A^0$
modes, the total contribution is dominated by boxes.

\begin{widetext}
\begin{figure}
\begin{center}
\includegraphics[width=7.8cm]{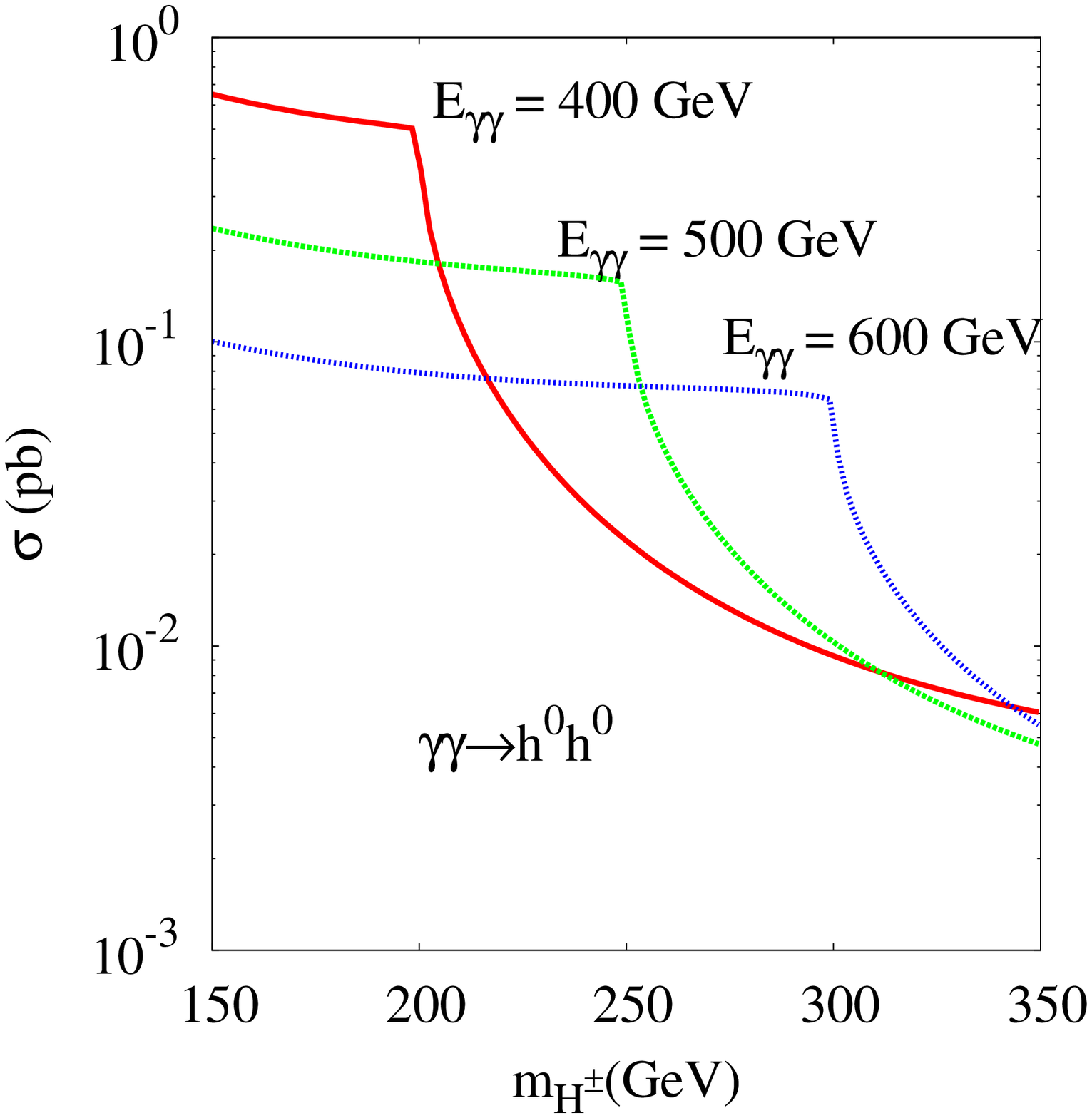}
\includegraphics[width=7.8cm]{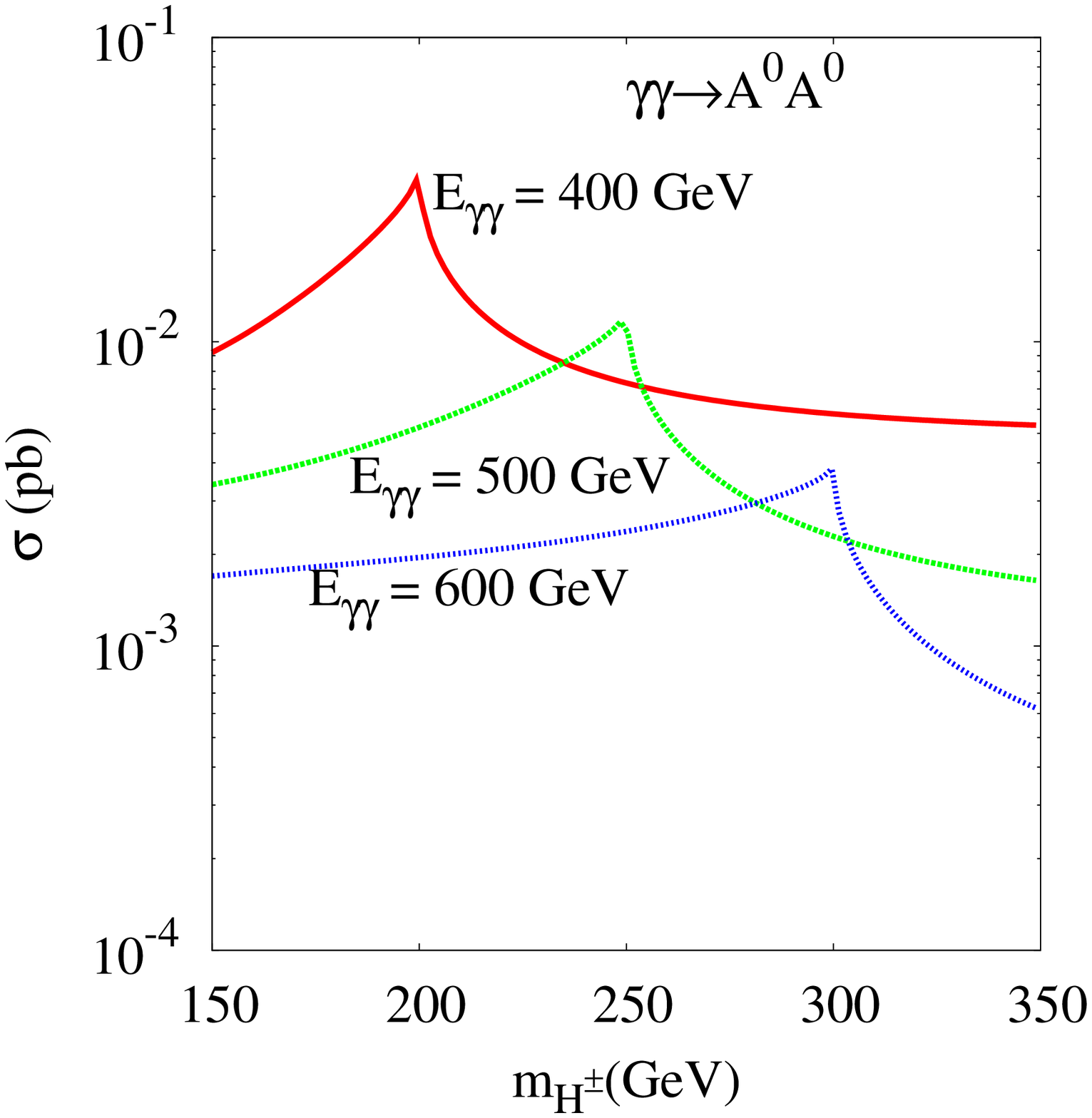}
\end{center}
\caption {The total cross section $\sigma(\gamma\gamma \to h^0h^0)$
(left) and $\sigma(\gamma\gamma \to A^0A^0)$ (right) as a function of
the charged Higgs mass for different center of mass energies $E_{\gamma\gamma}$ in the 2HDM.
With $ m_{h^0}, m_{H^0}, m_{A^0}, m_{12}  = 120, \,240, \,270, \, 350 \, \, GeV$,
$\sin\alpha = -0.9$ and $\tan\beta = 1$.}
\label{fig-mhp}
\end{figure}
\end{widetext}

In Fig.~\ref{fig-mhp}, we illustrate the sensitivity to the charged Higgs boson contribution for different center of mass energy for the $h^0h^0$ and the
$A^0A^0$ modes. In both cases, the cross sections are enhanced for light charged Higgs for the reasons explained above and are suppressed after crossing
the threshold for $\gamma\gamma\to H^+H^-$ production, i.e.,
$m_{H\pm}\ga E_{\gamma\gamma}/2$.

\begin{widetext}
\begin{figure}
\begin{center}
\includegraphics[width=7.5cm]{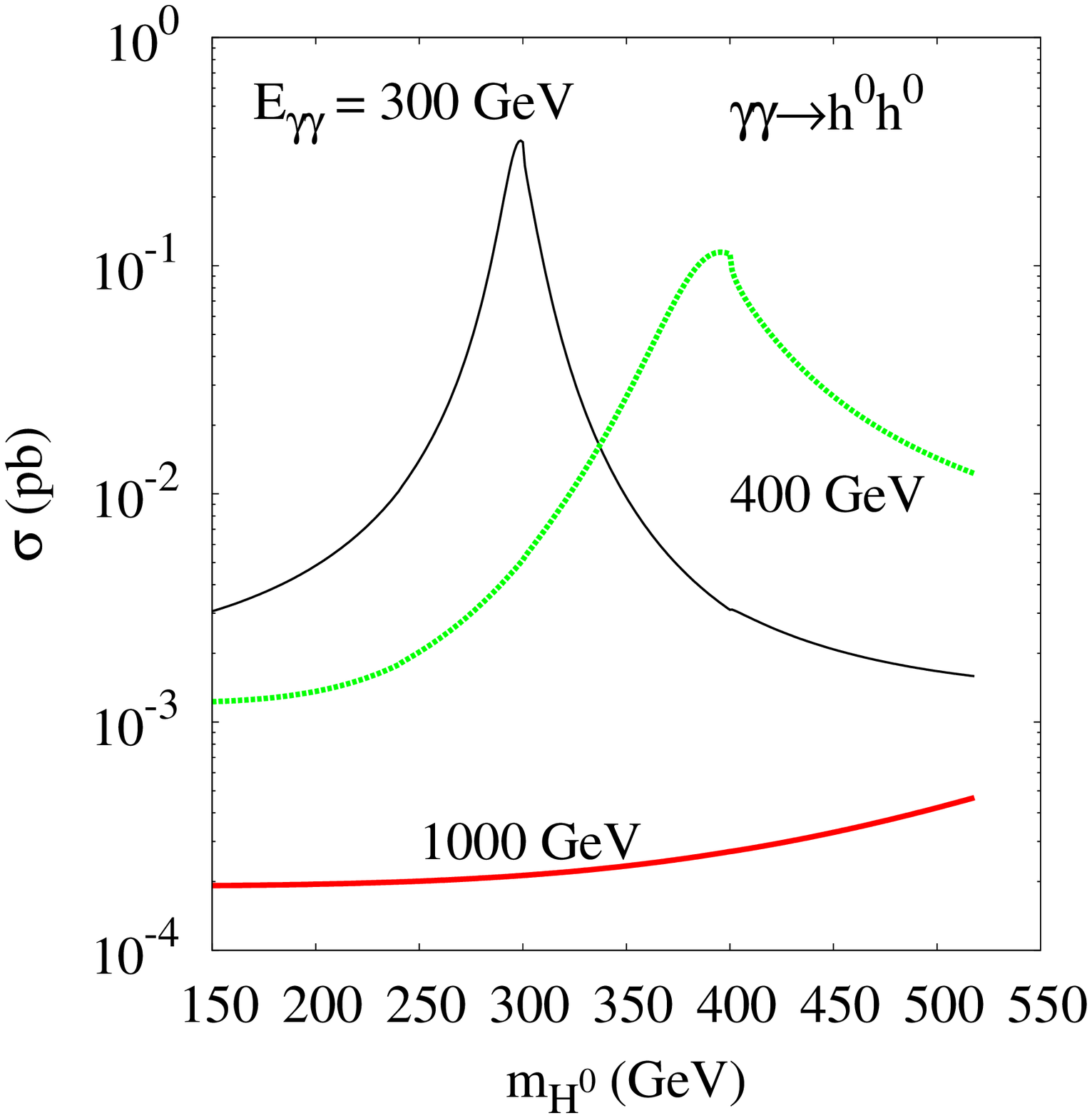}
\includegraphics[width=7.5cm]{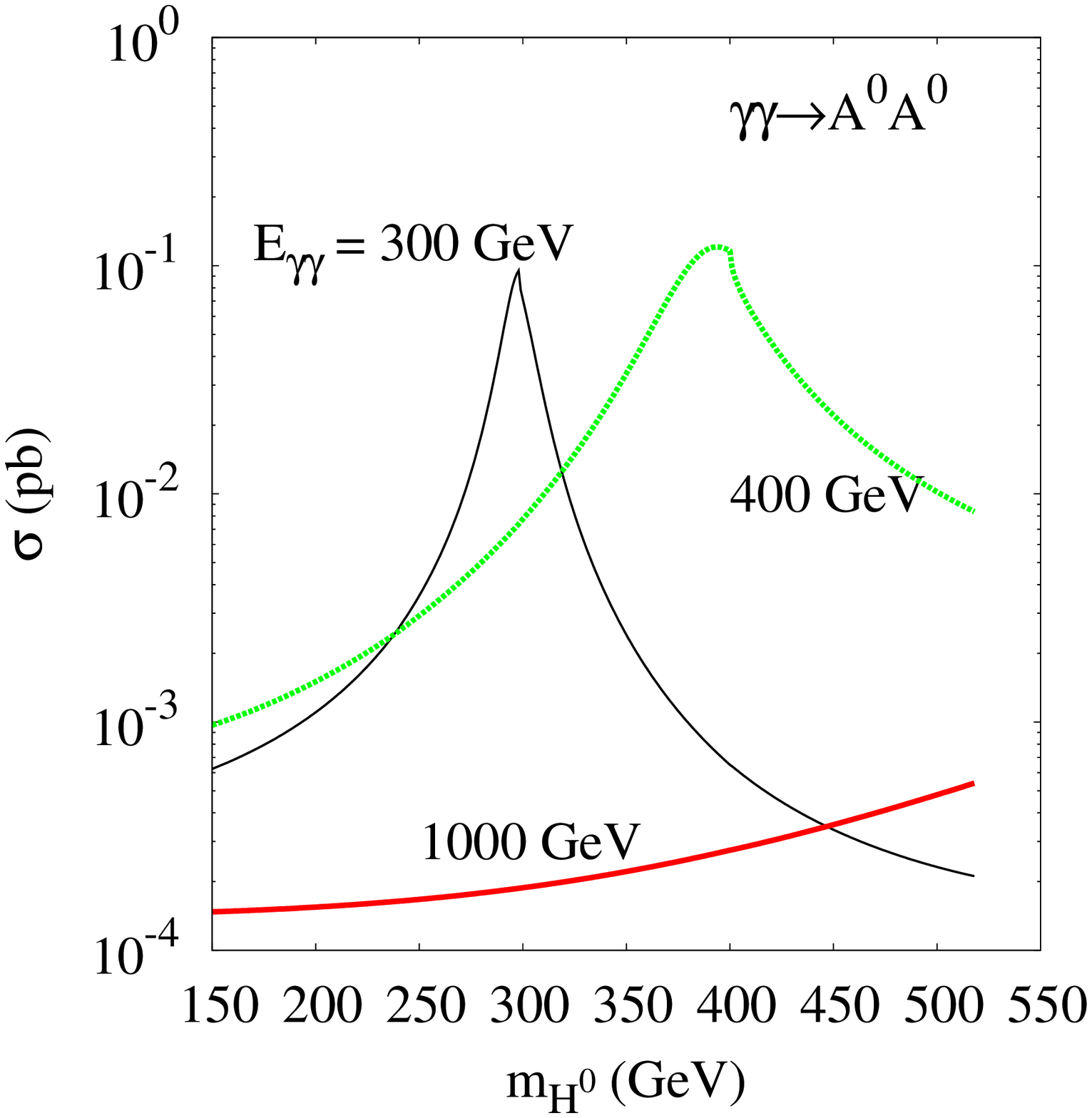}
\end{center}
\caption{The total cross section $\sigma(\gamma\gamma \to
h^0h^0)$ (left) and $\sigma(\gamma\gamma \to A^0A^0)$ (right) as a
function of the heavy Higgs mass $m_{H^0}$ in the 2HDM. With
$m_{h^0},m_{H^\pm} , m_{A^0}, m_{12} = \,120, \,250, \,150, \,200 \, 
\, GeV$, $\sin\alpha = 0.9$ and $\tan\beta = 1.5$ for different
values of the two photon center of mass energy $E_{\gamma\gamma}$.
} \label{fig3}
\end{figure}
\end{widetext}

In Fig.~\ref{fig3} we show the total cross section for
$\gamma\gamma \to h^0h^0$ (left) and $\gamma\gamma \to A^0A^0$
(right) as a function the heavy Higgs mass for several values of the center of mass energy. Once the center of mass energy is close to $m_{H^0}$, one can see in both plots
the effect of the resonance of the heavy CP-even Higgs.
The difference between them is only due to the vertex diagrams
with an intermediate heavy Higgs  that then decays to
$h^0h^0$ or $A^0A^0$.  In both cases, the cross sections
can reach 0.1 pb near the resonance $E_{\gamma\gamma}\approx m_{H^0}$.

\begin{widetext}
\begin{figure}
\begin{center}
\includegraphics[width=7.5cm]{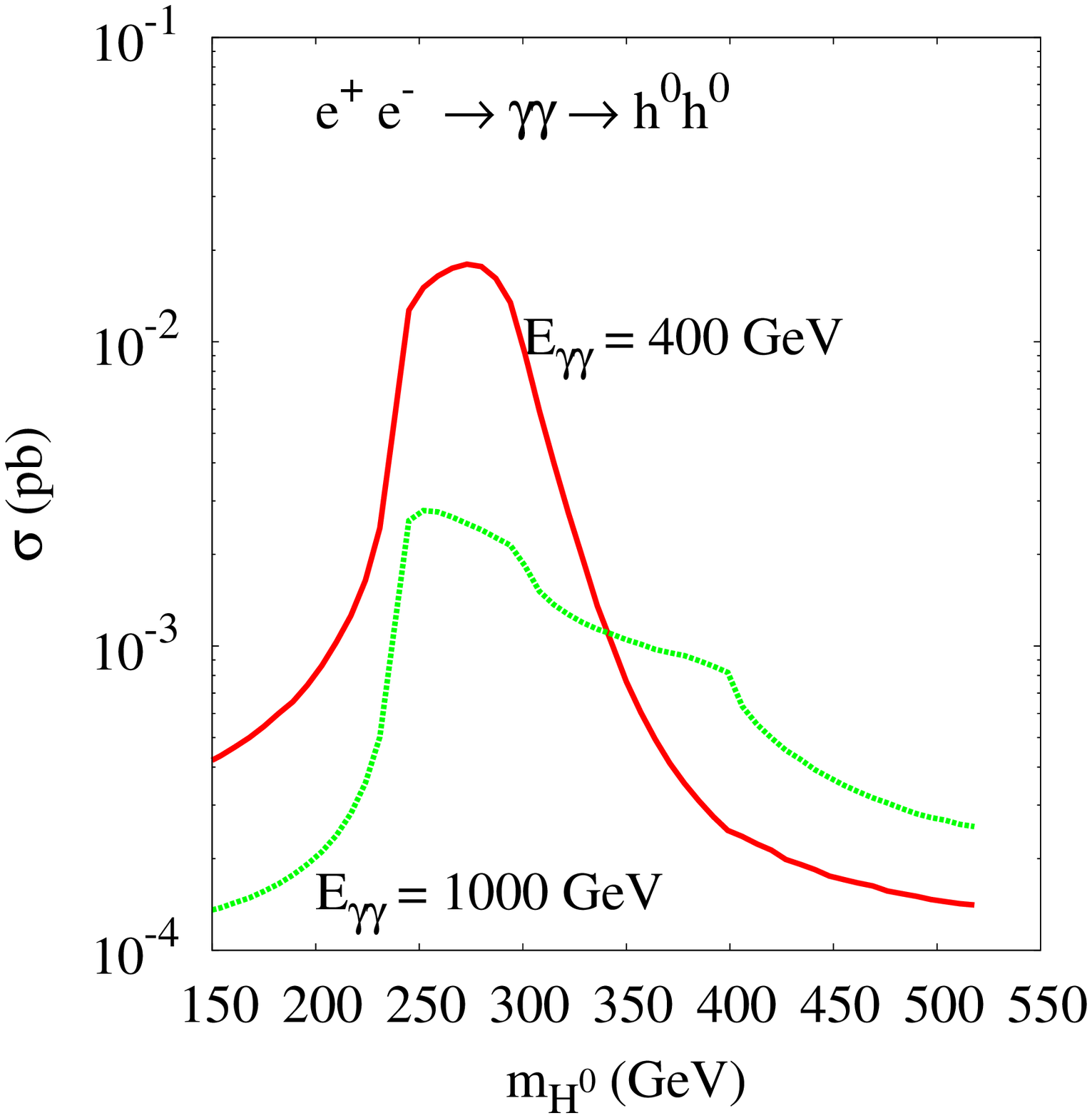}
\includegraphics[width=7.5cm]{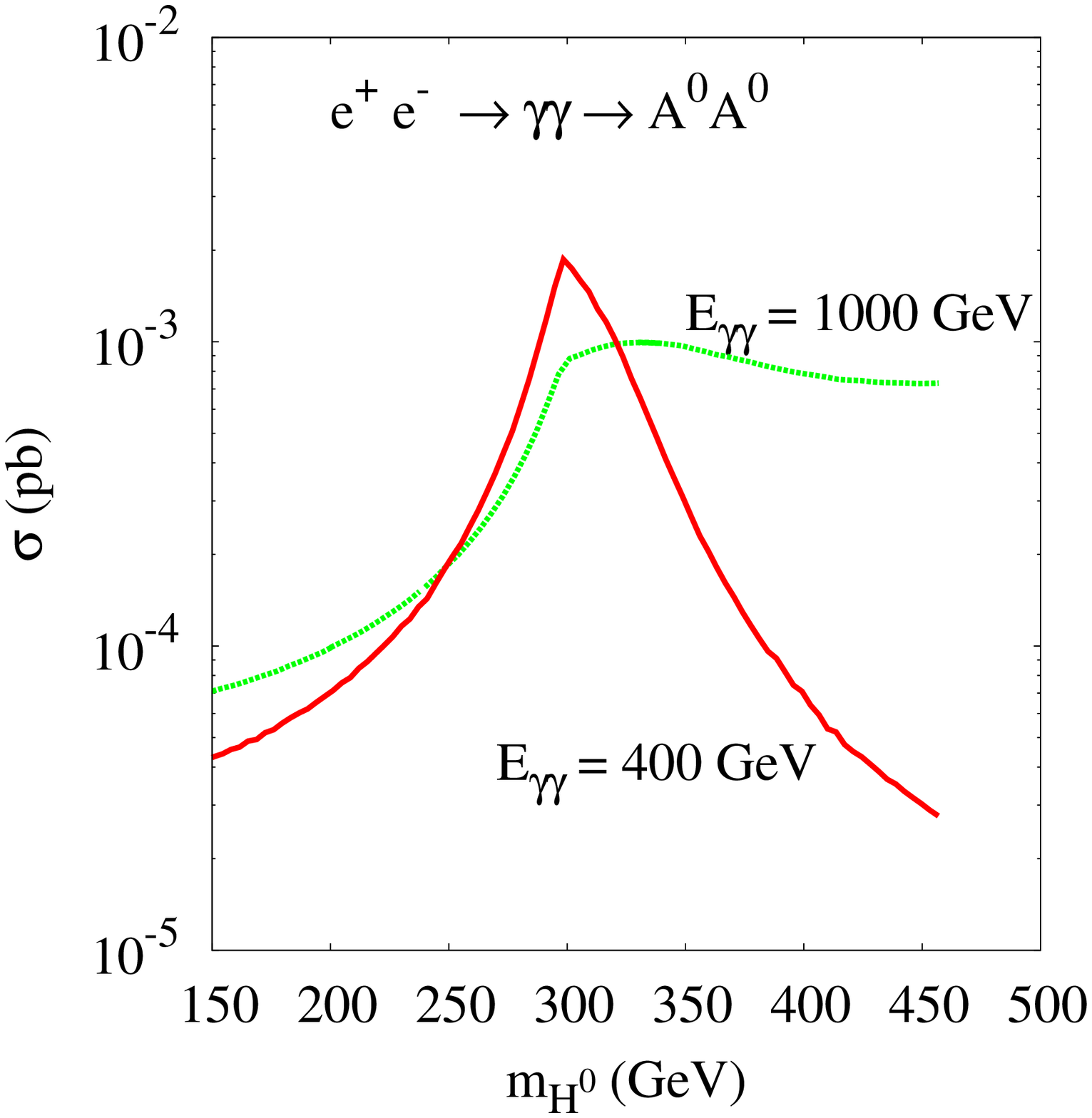}
\end{center}
\caption{The total cross section $\sigma(e^+ e^- \to
\gamma\gamma \to h^0h^0)$ (left) and $\sigma(e^+ e^- \to
\gamma\gamma \to A^0A^0)$ (right) as a function of the heavy Higgs
mass $m_{H^0}$ in the 2HDM with the same parameters as in Fig.~\ref{fig3}.}
\label{fig9}
\end{figure}
\end{widetext}

For the $h^0H^0$ or $H^0H^0$ final states, the generic set of Feynman diagrams that contribute to those processes is almost the same. The main enhancement factor for the cross section is the virtual charged Higgs bosons exchange, particularly relevant near the threshold region
$E_{\gamma\gamma}=2m_{H\pm}$. The only difference between  those final states and the $h^0 h^0$ and the $A^0 A^0$ ones, is the absence of the $H^0$ resonant effect, since it can not decay neither to $h^0H^0$ nor to $H^0H^0$. The situation is the same as in the SM. If the CP-even $H^0$ has a mass of the same order as $h^0$ the cross section of $h^0H^0$ or $H^0H^0$ could be of the same order of magnitude and may reach 0.1 pb. If the CP-even Higgs is heavy, phase space suppression occurs and the cross sections for $h^0H^0$ and $H^0H^0$ production are smaller.

In the case of $h^0A^0$ or $H^0A^0$ final state, a quite different situation occurs. Due to the presence of the CP odd scalar $A^0$ in the final state, the Higgs boson $\phi$ in the s-channel vertex (Fig.~\ref{fig:vert-diagrams}) $v_{1 \rightarrow4}$, $v_9$ and $v_{12}$ must also be CP-odd. Hence, the processes $\gamma \gamma \to h^0 A^0, H^0A^0$ do not proceed through an intermediate CP-even Higgs. This implies that there is no closed loop of virtual exchange of charged Higgs, both in the vertex and box contributions, which again, is the main factor of enhancement of the cross section for the $h^0h^0$ and the $A^0A^0$ modes. In the boxes, we have rather a mixture of charged Higgs and charged Goldstones in the loop (gauge couplings). We have checked that the values of the cross section of $\gamma\gamma\to h^0A^0$ are much smaller than the $h^0h^0$ and $A^0 A^0$ ones. We have performed a systematic scan over the 2HDM parameters with $m_{h^0}=m_{A^0}$ in the range 100 to 160 $GeV$, and found that the cross section of $\gamma\gamma\to h^0A^0$ does not exceed 0.04 $pb$. Note that the largest value for the cross section, 0.04 pb, is attained near the top quark threshold region $E_{\gamma\gamma}\approx 2m_t$; away from this region the cross section drops below 0.01 pb. Similarly, we found that large cross section for $\gamma\gamma\to h^0A^0$ prefer rather small values of $\tan\beta\la 3$, $|\sin\alpha| \ga 0.5$, large $m_{12}$ and also low center of mass energies $\sqrt{s}\la 600$ $GeV$.

Finally, in Fig.~\ref{fig9} we show the total cross section for $e^+ e^- \to \gamma\gamma \to h^0h^0$ (left) and for $e^+ e^- \to \gamma\gamma \to A^0A^0$ (right) as a function the heavy Higgs mass for two center of mass energies $E_{\gamma\gamma}$.
The total cross section is evaluated by convoluting the
photon-photon cross section with the photon-photon luminosity spectrum taken from the CompAZ library \cite{Zarnecki:2002qr}. CompAZ is based on formulae for the Compton scattering and provides the photon energy spectrum for different beam energies and the average photon polarization for a given photon energy.
First, let us remark that again this cross section is large enough to be measured in a significant region of the parameter space. We can still see the heavy Higgs resonance effects but somehow softened by the photon spectrum.
%
\begin{figure}[H]
\centering
\includegraphics[width=7.5cm]{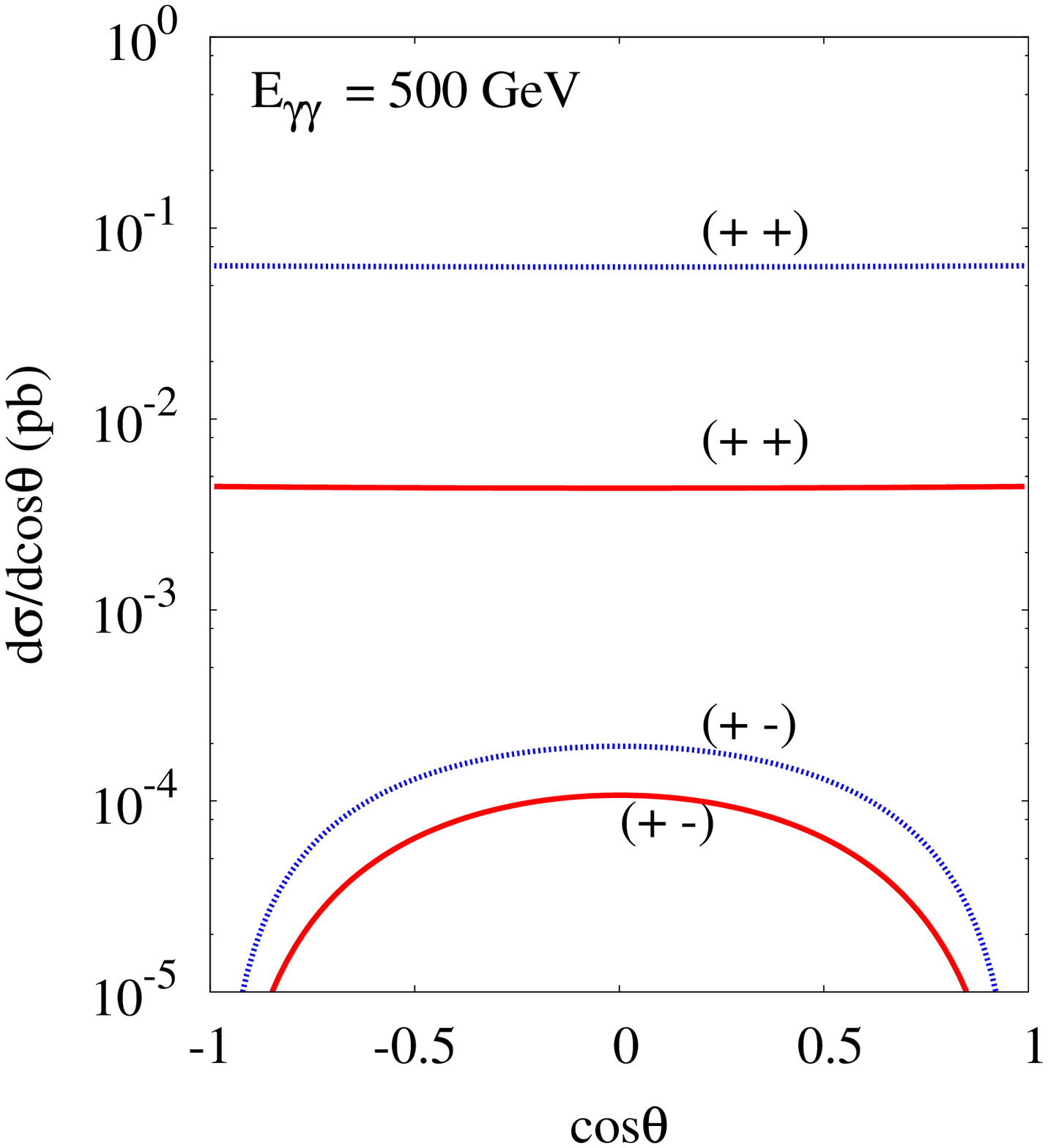}
\hskip 0.1cm
\includegraphics[width=7.5cm]{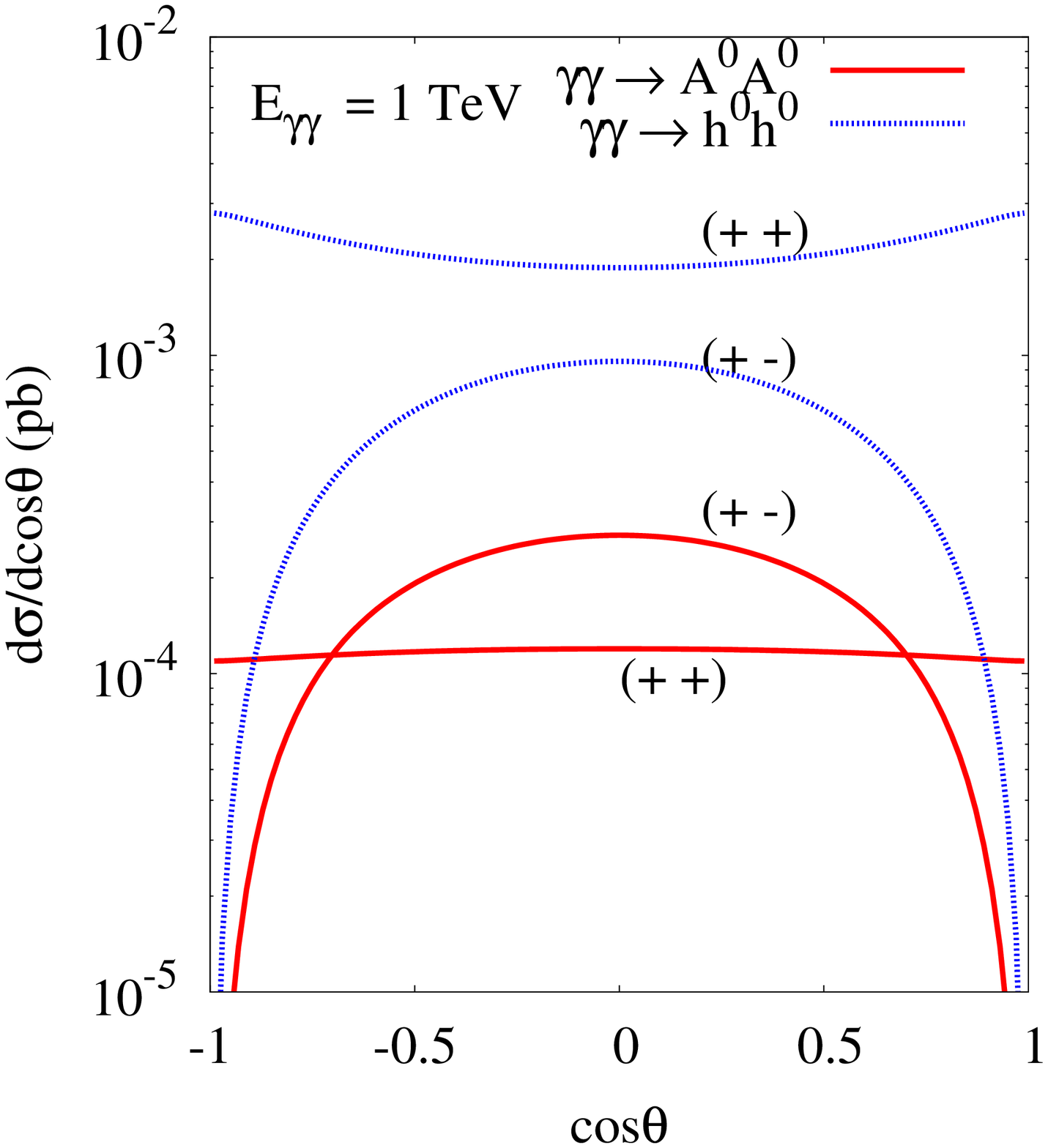}
\vskip -.3cm
\caption{The differential cross section for
$\gamma\gamma \to h^0h^0$ and $\gamma\gamma \to A^0A^0$
with equally polarized photons and oppositely polarized photons for $\sqrt{s} = 500$ GeV (left) and
$\sqrt{s} = 1$ TeV (right). The 2HDM parameters are the same as
in Fig.\ref{fig2} and $m_{H^\pm}$ fixed at 250 GeV. }
\label{ang}
\end{figure}
Let us now turn to the differential cross section for
$\gamma\gamma\to h^0h^0$ and $\gamma\gamma\to A^0A^0$. In
Fig.~\ref{ang} we illustrate the differential cross for a center
of mass energy of $\sqrt{s}=500$ $GeV$ and $\sqrt{s}=1$ $TeV$ and
for identically polarized $(++)$, $(--)$ and oppositely polarized
$(+-)$, $(-+)$ initial photons. As one can see from the left panel for
$\sqrt{s}=500$ $GeV$, for both the $h^0h^0$ and the $A^0A^0$
modes, the angular distribution for $(++)$ and $(--)$ polarized
photons is almost flat while for the $(+-)$ and $(-+)$ modes it
has a parabolic shape
 with a rather small cross section except for
 $\gamma\gamma\to h^0h^0$ where in the region of $-0.5<\cos\theta <0.5$
 the differential cross section is of the order of $0.1$ $fb$.
For $\sqrt{s}=1$ $TeV$, Fig.~\ref{ang} (right panel),
it is clear that the angular distribution are very similar to the
ones on the left panel with the cross sections more than one order
of magnitude smaller. The shape of these differential cross sections clearly tell us that we will not be able to distinguish the CP nature of the particle on the basis of this angular distribution.

\subsection{Fermiophobic limit}
In the SM, where just one doublet couples to all fermions, each
scalar couples to the different fermions with the same coupling
constant. In a general 2HDM it is also possible to couple just one
doublet to all fermions by choosing an appropriate symmetry for
both the fermions and the scalars. However, the difference between
the SM and the 2HDM is that now the couplings are proportional to
the rotation angles $\alpha$ and $\beta$. For instance, the
lightest CP-even Higgs couples to all fermions as
$\cos\alpha/\sin\beta$ $g^{SM}_{hf\bar{f}}$. By choosing $\alpha =
\pm \frac{\pi}{2}$, the lightest Higgs decouples from all
fermions. Such a scenario, with the appearance of the so-called
"fermiophobic" Higgs boson, arise in a variety of models
\cite{landsberg_cite}. The heavy CP-even scalar will acquire
larger couplings to the fermions than the corresponding SM
couplings. All the remaining scalars are not affected by this
choice as they do not couple proportionally to $\alpha$.
\begin{widetext}
\begin{figure}
\begin{center}
\includegraphics[height=3.7in,width=7.8cm]{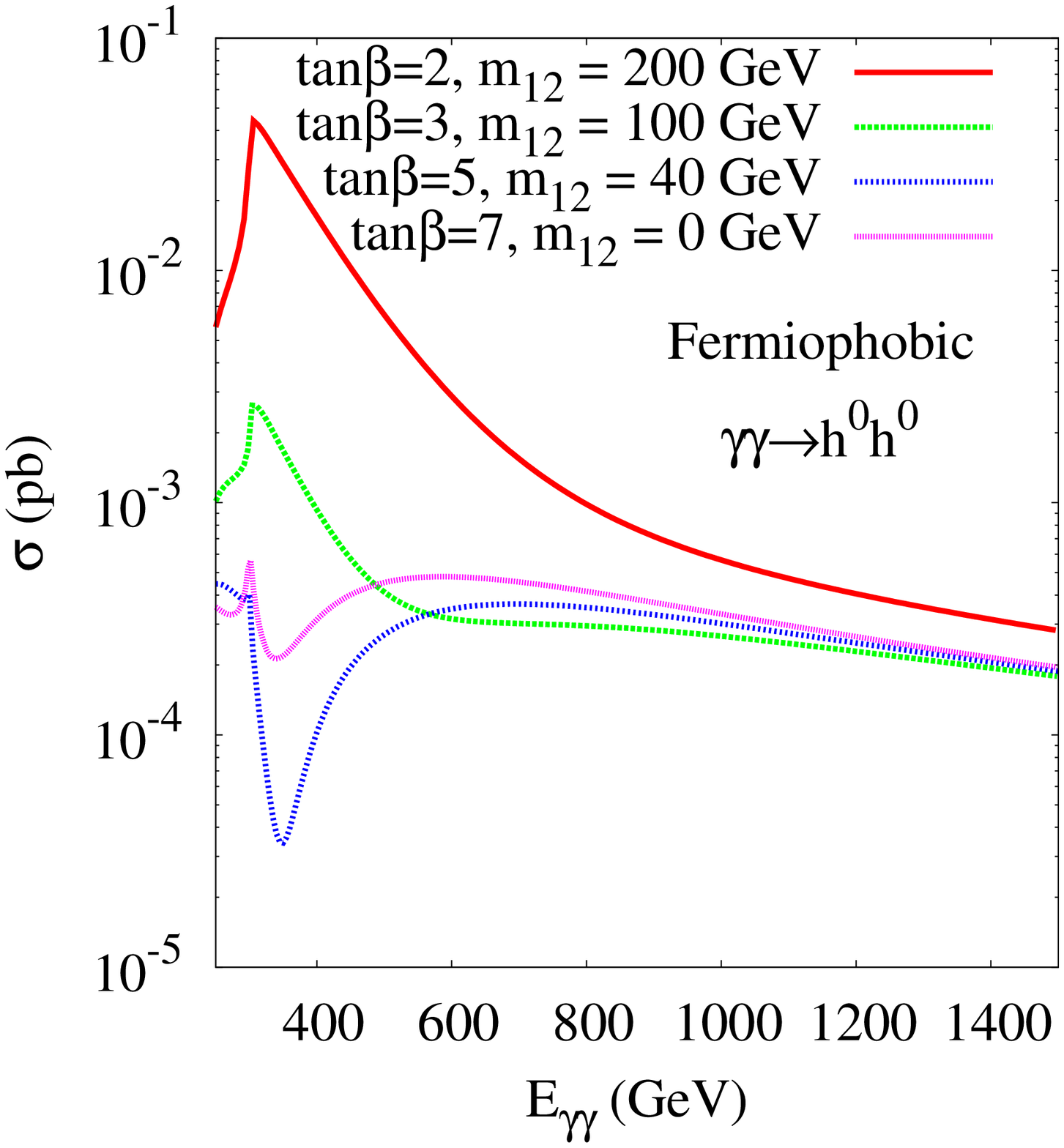}
\includegraphics[height=3.7in,width=7.8cm]{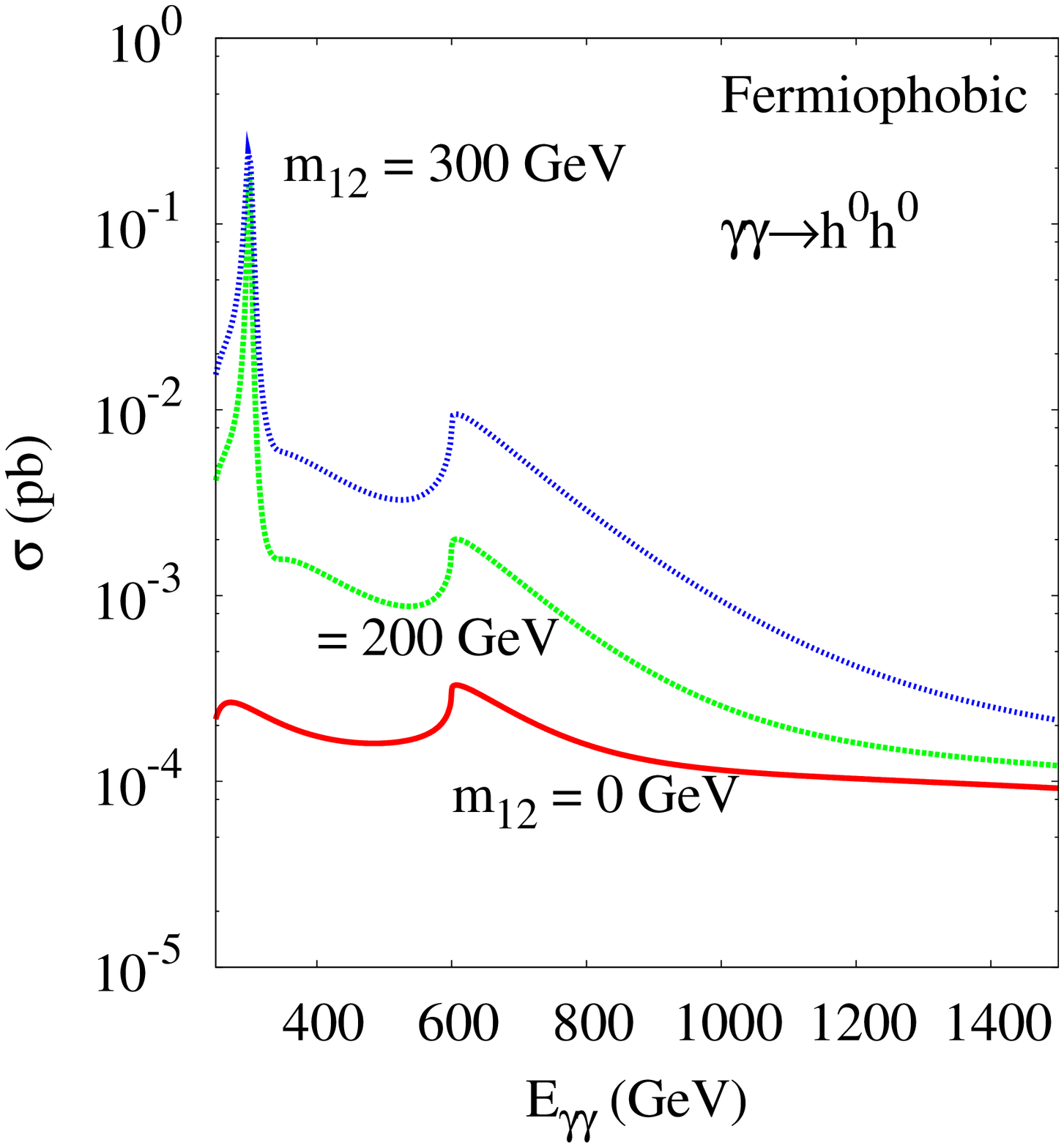}
\end{center}
\caption {The total cross section $\sigma(\gamma\gamma \to h^0h^0)$ as a
  function of the center of mass energy $E_{\gamma\gamma}$ in the
  fermiophobic limit ($\sin\alpha=1$).
In the left panel with different values of  $\tan\beta$
  and $m_{12}$ for $m_{h^0}, m_{H^0}, m_{A^0}, m_{H^\pm}$ =
100, 200, 100, 150 $GeV$.
In the right panel, $m_{h^0}=120$ $GeV$, $m_{H^0}=m_{H^\pm}=m_{A^0}$ = 300
 $GeV$ and $\tan\beta = 1$.}
\label{fig4}
\end{figure}
\end{widetext}

In the situation where the Higgs-fermion couplings are
substantially suppressed, the full decay width of the Higgs boson
is shared mostly between the $WW$, $ZZ$ and $\gamma\gamma$ decay
modes. In this limit, for masses $m_{h^0} < 100$ $GeV$, the Higgs
boson dominantly decays to photon pairs. Experimental searches for
fermiophobic Higgs bosons at the LEP collider and the Tevatron
collider have yielded negative results. Mass limits have been set
in a benchmark model that assumes that the couplings $hWW$ and
$hZZ$ have the same strength as in the SM and that all fermion
branching ratios are exactly zero. Combination of the results
obtained by the LEP collaborations \cite{aleph,delphi, l3,opal}
using the process $e^+ e^- \to h Z$, $h\to \gamma\gamma$ yielded
the lower bound $m_h > 109.7 $ $GeV$ at 95\%~C.L.~\cite{LEPlimit}.
In Run I, Tevatron has set lower limits on $m_{h}$ by the D0 and
CDF collaborations which are respectively 78.5 $GeV$
\cite{d0limit} and 82 $GeV$ \cite{cdflimit}, using the processes
$q\bar{q}^{\prime} \to V^* \to hW, hZ$, $h \to \gamma\gamma$, with
the dominant contribution coming from the $W$ boson. Recently the
CDF~\cite{CDFpage} and the D0~\cite{:2008it} collaborations have
improved their bounds which are now of the order of 100 $GeV$,
close to the bound obtained by the LEP collaborations. It should
be noted that all experimental mass bounds assume, in the
fermiophobic limit, $\tan \beta \sim 0$ in the tree-level
couplings to the gauge bosons.

It is clear that in the fermiophobic limit the coupling  $H^0 h^0
h^0$ (Eq.~(\ref{hll})) is directly proportional to $m_{12}$, while
the other couplings depend both on $m_{12}$, $\tan\beta$ as well
as on $m_{H^\pm}$.
We show in Fig.~\ref{fig4}, the
total cross section for $\gamma \gamma \to h^0 h^0$ as a function
of the two photon center of mass energy in the fermiophobic limit.
The parameters $m_{12}$ and $\tan\beta$ are varied within the allowed range.
It is clear that the cross section is
enhanced for large $m_{12}$ and can reach 0.1 pb for large $m_{12}=200$ GeV
(left) and $m_{12}=300$ GeV (right).
The observed kinks are the top threshold at $E_{\gamma\gamma}=2m_{t}$ and the charged Higgs threshold $E_{\gamma\gamma}=2m_{H\pm}$.

\subsection{Decoupling limit}

A study of 2HDM in the decoupling limit reveals the case where all
scalar masses except one formally become large and the effective
theory is just the SM with one doublet - $m_{h^0} << m_{\Phi}$ where
$m_{\Phi} = m_{H^0, A^0, H^\pm}$ (see \cite{Gunion:2002zf} for an overview).
In this case, the CP-even $h^0$ is the lightest scalar particle while the
other Higgs particles $H^0$, $A^0$ and $H^\pm$ are extremely heavy.
In 2HDM, the decoupling limit can be achieved by taking the limit
$\alpha \to \beta - \pi/2$. This means that the coupling of the $h^0$
to the gauge bosons, fermions and light Higgs, $h^0$ are the same as
for the Standard Model $h^{SM}$ Higgs. Also, in the decoupling limit,
the triple Higgs coupling $\lambda^{(0)}_{h^0h^0 H^0}$  vanishes at
tree-level, so
that the heavy Higgs cannot contribute to the process $\gamma
\gamma \to h^0 h^0$ and the result is independent of the mass
$m_{H^0}$. In the decoupling limit, the tree-level trilinear Higgs
couplings take the form
\begin{eqnarray}
\lambda^{2HDM}_{h^0h^0h^0} &\approx& \lambda^{SM}_{h^0h^0h^0},\\
\lambda^{2HDM}_{h^0h^0H^0} &\approx& 0,\nonumber\\
\lambda^{2HDM}_{h^0H^+H^-} &\approx& -\frac{g}{2 m_W}
\bigg[ m^2_{h^0} + 2 m^2_{H^\pm} + m^2_{12} \bigg],\nonumber\\
\lambda^{2HDM}_{h^0h^0H^+H^-} &\approx& \frac{g}{2 m_W}
\lambda^{2HDM}_{h^0H^+H^-}. \label{DL}
\end{eqnarray}
It is clear that these couplings are independent of $\tan\beta$ as well.
As one can see from the analytical expression of
$\lambda^{2HDM}_{h^0H^+H^-}$
and $\lambda^{2HDM}_{h^0h^0H^+H^-}$, the charged Higgs and $m_{12}^2$
add constructively for $m_{12}^2>0$ and destructively for
 $m_{12}^2<0$.
At tree level, the enhancement of the cross section essentially depends
on the size of the $h^0 H^+ H^-$ and $H^0 H^- H^+$ couplings. By taking
these couplings as large as possible under the referred theoretical and
experimental constraints, we obtain the best possible enhancement for the cross section $\sigma (\gamma \gamma \to h^0 h^0 )$ in the 2HDM for each
$m_{\Phi}$ ($m_{\Phi} = m_{H^0} = m_{A^0} = m_{H^\pm}$).
However, it is well known that those couplings of the CP-even
$h^0$ which mimic the SM couplings get significant radiative
corrections in the decoupling limit to which we refer to as
non-decoupling effects. Several studies have been carried out
looking for non-decoupling effects in Higgs boson decays and Higgs
self-interactions.  Large loop effects in $h^0\to \gamma \gamma$,
$h^0\to \gamma Z$ and $h^0\to b\bar{b}$ have been pointed out for the
2HDM~\cite{maria,indirect1} and may provide indirect information on
the Higgs masses and the involved triple Higgs couplings such as
$\lambda_{h^0H^+H^-}$ for $h^0\to \gamma \gamma$ and
 $\lambda_{h^0H^0H^0}$, $\lambda_{h^0A^0A^0}$ and
$\lambda_{h^0h^0h^0}$ for $h^0\to b\bar{b}$.

The non-decoupling contributions to the triple Higgs self-coupling
$\lambda_{h^0h^0h^0}$ have been investigated in the 2HDM in
Ref.~\cite{okada}, using the Feynman diagrammatic method. It has been
demonstrated that the one-loop leading contributions originated from
the heavy Higgs boson loops and the top quark loops to the effective
$h^0h^0h^0$ coupling can be written as ~\cite{okada}
\begin{eqnarray}
 \lambda_{h^0h^0h^0}^{eff} \!\!&=&\!\! \frac{3 m_{h^0}^2}{v}
      \left\{ 1
     + \frac{m_{H^0}^4}{12 \pi^2 m_{h^0}^2 v^2}
     \left(1 + \frac{M^2}{m_{H^0}^2}\right)^3
     + \frac{m_{A^0}^4}{12 \pi^2 m_{h^0}^2 v^2}
     \left(1 + \frac{M^2}{m_{A^0}^2}\right)^3 \right.\nonumber\\
&&\left. \!\!\!\!\!\!\!\!
    + \frac{m_{H^\pm}^4}{6 \pi^2 m_{h^0}^2 v^2}
    \left(1 + \frac{M^2}{m_{H^\pm}^2}\right)^3
     - \frac{N_c m_t^4}{3 \pi^2 m_{h^0}^2 v^2} +
      {\cal O} \left(\frac{p^2_i m_\Phi^2}{m_{h^0}^2 v^2},
     \;\frac{m_\Phi^2}{v^2},
     \;\frac{p^2_i m_t^2}{m_{h^0}^2 v^2},
     \;\frac{m_t^2}{v^2}  \right)
      \right\} ~,
\label{ceff}
\end{eqnarray}
where $M^2=m_{12}^2/(\sin\be\cos\be)$, $m_\Phi^{}$ and $p_i$
represent the mass of $H^0$, $A^0$ or $H^\pm$ and the momenta of
external Higgs lines, respectively, $N_c$ denotes the number of
colors, and $m_t$ is the mass of top quark.  We note that in
Eq.~(\ref{ceff}) $m_{h^0}$ is the renormalized physical mass of
the lightest CP-even Higgs boson $h^0$. In the calculation of
the $\gamma \gamma \to h^0h^0$ cross section in the decoupling
limit, we replace the $\lambda^{(0)}_{h^0h^0h^0}$ coupling by its
effective coupling given in Eq.~(\ref{ceff}) which corresponds,
in this limit, to an effective 2-loop 2HDM contribution
(see Ref.~\cite{okada} for a detailed discussion).
\begin{figure}[H]
\centering
\includegraphics[width=7.5cm]{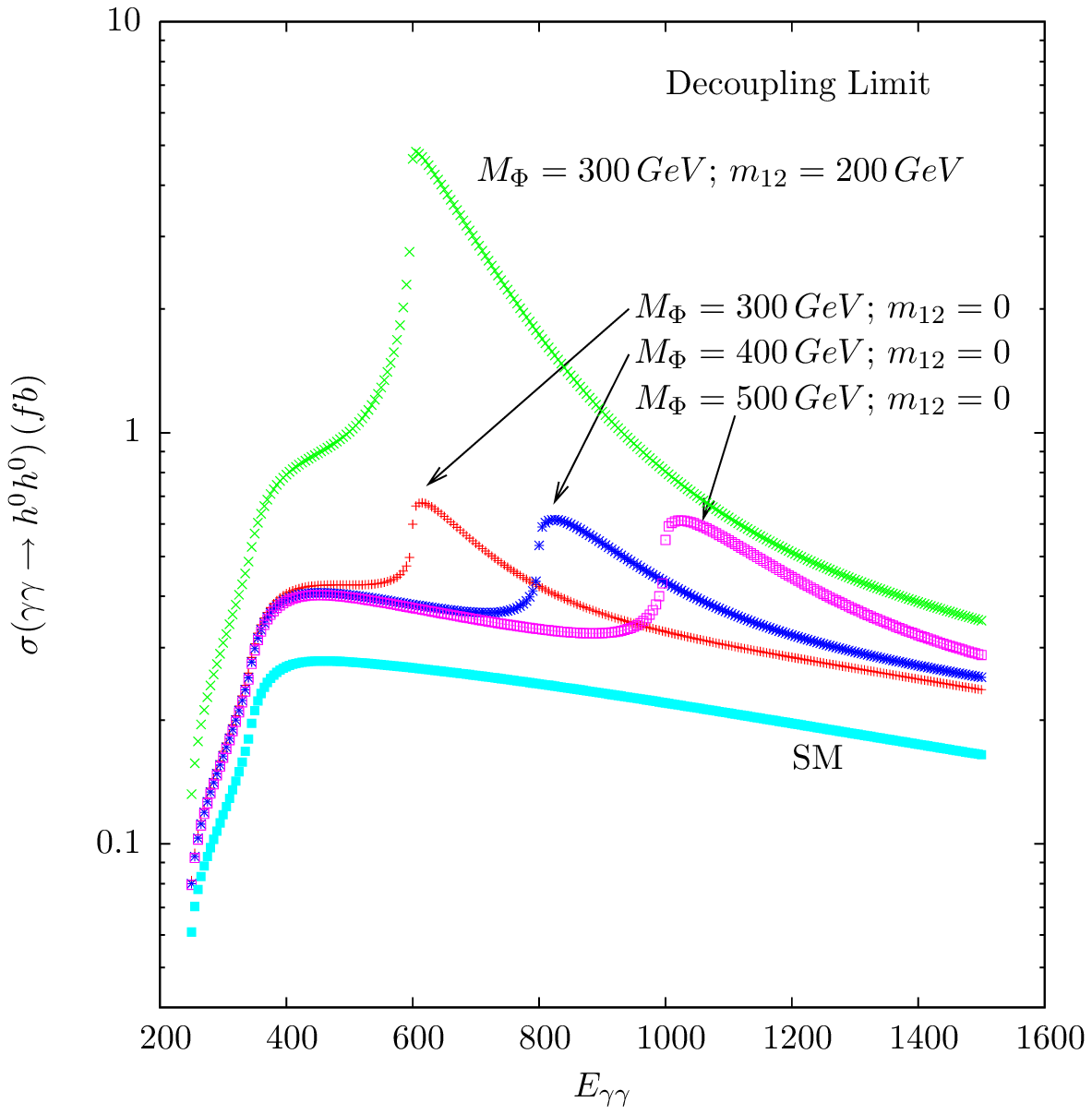}
\hskip 0.1cm
\includegraphics[width=8.7cm]{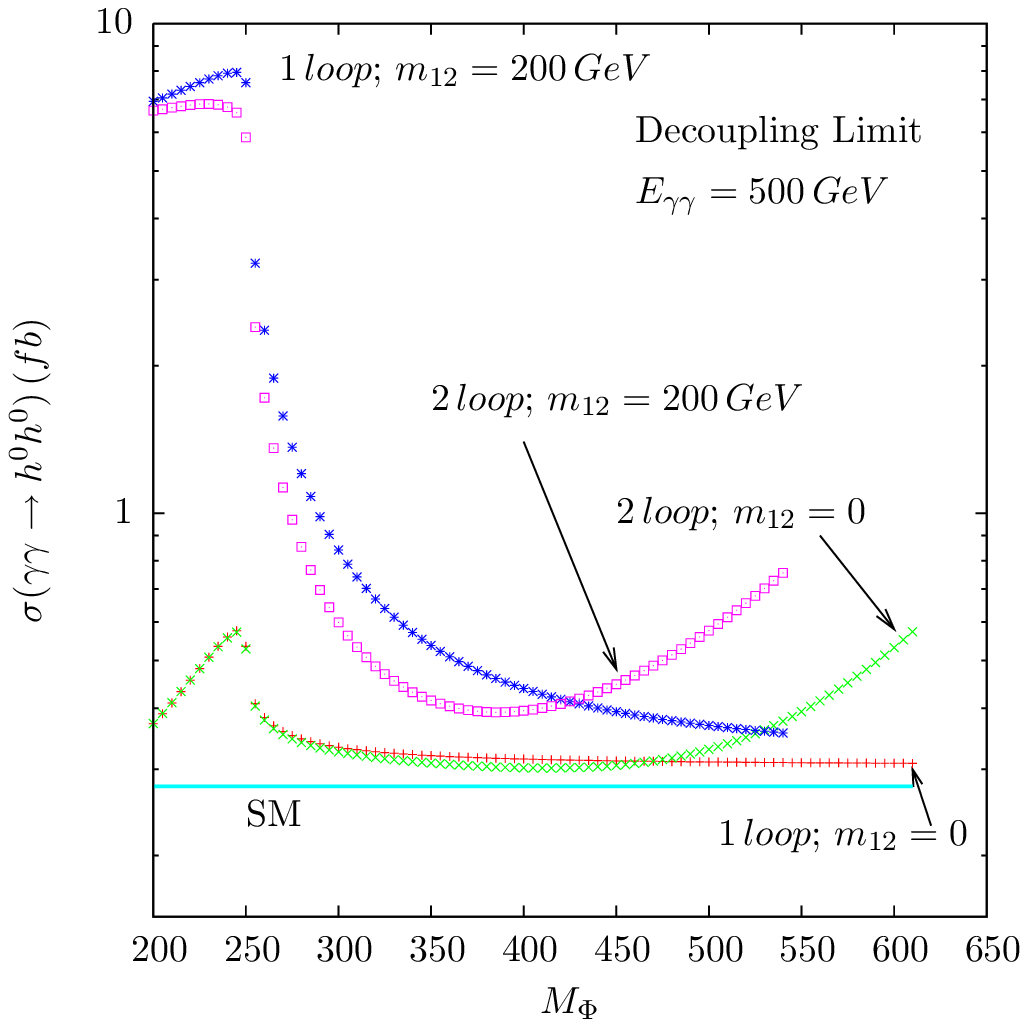}
\vskip -0.5cm \caption {Cross sections for $h^0 h^0$ production in the
decoupling limit with unpolarized photons. On the right we show the
loop contributions to the total cross section as a
function of $ m_{\Phi}$ and for two values of $m_{12}$, 0 and 200 GeV.
On the left panel the cross section as a
function $E_{\gamma \gamma}$ is shown for different values of
$m_{\Phi}$ and $m_{12}$. The light Higgs mass is $m_{h^0} = 120$ $GeV$.}
\label{figdl}
\end{figure}

As stated in the introduction, this process was studied in detail in
Refs.~\cite{hollik:2008, knemaura} in the decoupling limit.
In fact, there is a one to one correspondence between our potential and the
 potential used in \cite{hollik:2008} which relate
$\lambda_5$ parameter of Ref~\cite{hollik:2008} to our $m_{12}$ by:
\begin{eqnarray}
m_{12}^2=-\lambda_5 v_1 v_2 \label{1to1}
\end{eqnarray}
From this relation, one can see that $\lambda_5 < 0$ (resp $\lambda_5 > 0$)
correspond to  our $m_{12}^2>0$ (resp $m_{12}^2<0$).
In \cite{hollik:2008}, the non-decoupling effects have their origin in the
$h^0H^+H^-$ and $h^0h^0H^+H^-$ vertex and are present already at tree level.
In their notation, when $\lambda_5 =0$ ($m_{12}=0$ in our notation) and at
the same time the charged Higgs mass and the lightest Higgs mass are of
the same order we get $\lambda_{h^0H^+H^-} \approx \lambda^{SM}_{h^0h^0h^0}$.
Hence, in this limit the 2HDM cross section is very similar to its SM
counterpart. As $m_{12}^2$ falls to negative values, the cross
section drops due to a cancelation
between the $m_{12}^2$ and the $m_{H^\pm}^2$ contributions until the $m_{12}^2$
term starts to dominate and the cross section increase again. For positive
$m_{12}^2$ the cross section is enhanced relative to the SM one and the
non-decoupling effects can be quite large as shown in \cite{hollik:2008} for
$\lambda_5<0$.
Therefore, at the 1-loop level $m_{12}^2$ and the charged Higgs mass are the
parameters that regulate the non-decoupling effects. In ~\cite{knemaura},
the authors have studied only the case of $m_{12}=0$ and
looked for non-decoupling effects in higher order
corrections with origin in the vertex $h^0 h^0 h^0$ as described earlier.
In this case the non-decoupling effects appear mainly for large values of
the masses. In this section we will combine both the effects of
ref~\cite{hollik:2008} and ref~\cite{knemaura} and present
the results for the case of the unpolarized photon cross section. We will
show that even in this case where the cross sections would be severely
reduced, there are still regions where the 2HDM CP even Higgs
$h^0$ could be disentangled from the SM $h_{SM}$.

In the left panel of Fig.~\ref{figdl} we show the cross section for 
$\gamma \gamma \to h^0 h^0$ as a function of $E_{\gamma\gamma}$ for
$m_{\Phi} =$ 300, 400 and 500 $GeV$ with $m_{12}=0$ together with the
case where $m_{12}=200$ $GeV$ (note that from now on we will be
considering $m_{12}^2>0$)
and $m_{\Phi} = 300$ $GeV$.
In this left panel, the coupling $h^0h^0h^0$ is taken at
the tree-level without the higher order correction of Eq.(\ref{ceff}).
The non-decoupling effects due to the charged Higgs mass can be seen in
the vicinity of $E_{\gamma\gamma}$ = 2 $m_{\Phi}$ where the cross section
is of the order of 0.6 to 0.7 fbarn. This effect is enhanced for higher
values of $m_{12}$. As shown in the plot,
for a charged Higgs mass of 300  $GeV$ and $m_{12}=200 \, GeV$
the cross section can reach 5 fbarn and this
value grows with large $m_{12}$.
In the right panel of Fig.~\ref{figdl} we display the cross
section of $\gamma \gamma \to h^0 h^0$ as a function $m_{\Phi}$. Here,
besides the SM value we plot four different scenarios. The one-loop case
with $m_{12}=0$ and $m_{12}=200 \, GeV$ and the two-loop case with the
higher order corrections given in Eq.(\ref{ceff}) for the same two
values of $m_{12}$.
One can see that the cross section enhancement due to the large corrections
in $\lambda^{eff}_{h^0h^0h^0}$ take place only for large $m_{\Phi}$ if
$m_{12}=0$. As $m_{12}$ grows the cross section grows as described in the left panel, but on top of that we get an extra enhancement due to the higher order corrections. Largest values of the cross section,
that can reach 10 fbarn, are attained for the low  mass region in
$m_{\Phi}$. It is also clear from the right panel that without the higher
order corrections of Eq.(\ref{ceff}), for large $m_{\Phi}$, the 2HDM cross
section is similar to SM one. On the other hand, once those higher order
corrections are included, non decoupling effects arise - the
cross section then grows relative to the SM one and can be almost two
orders of magnitude above the corresponding SM process depending on
the value of $m_{\Phi}$. We finally note that the cut on $m_{\Phi}$ at 610
$GeV$ for $m_{12}=0$ and on $m_{\Phi}$ at 550
$GeV$ for $m_{12}=200\ GeV$ is due to unitarity constraints.

\section{Higgs signatures}
\label{sec:signatures} We are considering a light CP-even Higgs,
that is, with a mass of 120 $GeV$ or less. Assuming that all decay
channels with some other Higgs boson in the final state are
unaccessible, this particle decays predominantly to $b \bar{b}$ in
this mass region. The exception is in the fermiophobic Higgs
scenario where it decays to two photons although for a mass of 120
$GeV$ one has already to consider the decay to two W bosons even
if one of the W is off-shell and strongly virtual. The rate at
which it decays to each final state depends on the remaining
parameters of the 2HDM (see~\cite{santos,Kanemura:2009mk} for details).
The two subleading decays that compete with $h^0 \rightarrow b \bar{b}$
are $h^0 \rightarrow c \bar{c}$ and $h^0 \rightarrow \tau^+
\tau^-$. In model type I the branching fractions to fermions are
the SM ones because the coupling dependence cancels. In model type
II, the ratio $\Gamma (h^0 \rightarrow b \bar{b})/\Gamma (h^0
\rightarrow \tau^+ \tau^-)$ is the SM one. On the other hand it is
easy to check that for $\tan \beta \geq 1$ the decay $h^0
\rightarrow b \bar{b}$ is again the dominant one provided we are
in a region with moderate values of $\tan \alpha$. For the
remaining Yukawa models the situation does not change
dramatically. However, if we take as an example the case where
$\Phi_2$ couples to the quarks and $\Phi_1$ couples to the
leptons, we obtain the ratio
\begin{equation}
\frac{\Gamma (h^0 \rightarrow b \bar{b})}{\Gamma (h^0 \rightarrow
\tau^+ \tau^-)} \, = \, \frac{m_b^2}{m_{\tau}^2} \,
\frac{1}{\tan^2 \alpha \, \tan^2 \beta}
\end{equation}
in the limit $m_h \gg m_q$. Even for $\tan \alpha = \tan \beta
\approx 3$ the ratio becomes almost 100 times smaller than the
corresponding SM ratio. Therefore, a detailed study for each model
will have to take into account the exact branching fractions for
each Yukawa version of the 2HDM.

The dominant background to double Higgs production is $\gamma
\gamma \rightarrow W^+ W^-$ and non-resonant four jet production.
The first one can be reduced by imposing a cut
on the invariant mass of each pair of b-jets, $M(q \bar{q})$,
forcing it to be close to Higgs mass. An efficient b-tagging would
further reduce the background by asking that at least three jets
be identified as originating from $b$ quarks. A cut on the polar
angle would eliminate the non-resonant 4-jet background. Together they would
reduce the backgrounds to a level well below the signal. A more
detailed analysis can be found in \cite{Belusevic:2004pz}. In the
fermiophobic case, the analysis is greatly simplified by the
smallness of the four photon production cross section. We just
need to avoid very soft photons which can be done with a sensible
cut on the photon's transverse momentum.

The process $\gamma \gamma \rightarrow h^0 H^0$ gives rise to very
similar signatures. There are in principle two drawbacks: first
the phase space is reduced because there is a heavier Higgs in the
final state; second we can not reduce the background by asking the
two invariant masses from each pair of b-jets to have a similar
magnitude. On the other hand, if the channel $H^0 \rightarrow
h^0h^0$ is open, it can be dominant. This would lead to a very
interesting signal of a six b-jet final state. As for $\gamma
\gamma \rightarrow h^0 (H^0) A^0$ we concluded that the cross section is much smaller due to the absence of virtual charged Higgs in the loop.

A pseudo-scalar Higgs decays again mainly as $A^0 \rightarrow b
\bar{b}$ and the subleading competing decays are again $A^0
\rightarrow c \bar{c}$ and $A^0 \rightarrow \tau^+ \tau^-$. The
situation is similar to the CP-even Higgs case - if $\tan \beta
\approx 1$ the $b \bar{b}$ channel is always the dominant and well
above the others. Once more if we consider the model where
one doublet couples to the quarks and the other couples to the
leptons, we obtain the ratio
\begin{equation}
\frac{\Gamma (A^0 \rightarrow b \bar{b})}{\Gamma (A^0 \rightarrow
\tau^+ \tau^-)} \, = \, \frac{m_b^2}{m_{\tau}^2} \, \frac{1}{
\tan^2 \beta}
\end{equation}
in the limit $m_h \gg m_q$. In this case, for $\tan \beta \approx
3$ the ratio becomes 10 times smaller than the corresponding SM
ratio. Note that a similar situation can occur in the ratio
between decays to down and to up quarks.
Obviously, this is true as long as the other channels involving
other Higgs are closed. The decays to either $h^0 Z $ or $W^+ H^-$
become dominant as soon as they are kinematically allowed. The
last two cases could again lead to interesting final states which
are easy to detect. As all the cases discussed refer to light Higgs,
even if the Higgs to Higgs channel is open the Higgs decay width
will in most situation be well below the $GeV$ or a few $GeV$ at most.

A final word about the behavior of the cross section with the
scattering angle. We have shown that if the $A^0$ and $h^0$ masses
are of the same order, and because in most models the possible
final states are very similar, it will be very hard to distinguish
a CP-even from a CP-odd state. In fact, even if one changes the
polarization of the initial photons, the differential cross
section does not distinguish clearly between the two cases except
in regions where either the cross sections are too small to be
measured or the angle is too small to be probed.

\section{Conclusions}
\label{sec:summary}
We have calculated the total cross section for
$\gamma \gamma \to S_iS_j$, $S_i=h^0, A^0, H^0$, in the framework of the
2HDM taking into account perturbativity, unitarity as well as vacuum
stability constraints on the scalar potential parameters $\lambda$'s.
All available experimental constraints were also taken into account.
For the numerical study, we mainly focused on the
$h^0h^0$ and $A^0A^0$ production modes. We have studied those
processes in the general 2HDM, and in two other
limiting scenarios: the fermiophobic limit, a scenario where the
lightest CP-even Higgs decouples from the fermions and the
decoupling scenario where this same Higgs resembles the SM Higgs boson.
We have shown that, for both production modes, the most important
contribution to $\sigma (\gamma \gamma \to h^0h^0)$ and to
$\sigma (\gamma \gamma \to A^0A^0) $ comes from the charged Higgs
$H^\pm$ diagrams and also from the diagrams with a resonant heavy Higgs
that can decay as $H^0\to h^0h^0$ or $H^0\to A^0A^0$.
Since the cross section is dominated by the charged Higgs virtual exchange, it does not depend on the Yukawa structure of the model.

We have shown that the cross section for $\gamma \gamma \to h^0h^0$
can be more than 100 times larger than the corresponding SM one in vast regions
of the parameters space.
The parameter space will easily be probed for the largest allowed
values of $\tan \beta$, $m_{12}^2$ and $|\sin \alpha|$.
A light charged Higgs, that is, below the collider center of mass energy,
is preferred. A variable energy collider would be a good option to
detect the heavy Higgs resonance.
We have argued that knowledge of the charged Higgs effects may be crucial
to understand the nature of the Higgs bosons if they are eventually found
in future experiments at LHC and/or ILC. In the case of
$\gamma\gamma\to h^0h^0$, we have illustrated that even with
a charged Higgs mass of the order of 300 GeV, in agreement with
$b\to s \gamma$, the cross section can have a substantial enhancement (clearly at least one order of magnitude above the SM). In case of 2HDM type I, where a light charged Higgs is
not ruled out from $b\to s \gamma$ one can have an even larger
enhancement which could be 3 order of magnitude above the SM results.

The analysis in \cite{Belusevic:2004pz} shows that the SM Higgs
triple coupling could be probed at a linear collider. As described
before, their analysis is mainly based on an invariant mass cut, on the
identification of at least 3 jets as originating from b-quarks and on
a the polar angle cut $|\cos \theta_b| < 0.9$. We have showed that the
inclusion of the new 2HDM diagrams do not change the angular
distribution as the Higgs angular distribution remains almost flat,
so that the same cut could be applied. Moreover, in the two Yukawa
versions of the model, $BR(h \to b \bar{b})$ is at least the SM one
if not larger. Because the invariant mass cut is the same, the
analysis can be applied directly to the 2HDM case. Therefore,
when a complete experimental analysis is completed for the SM, it is ready
to be used to constraint the 2HDM parameter space. This is one
of the major advantages of this study.
%

Although other regions give rise to higher cross sections,
the very interesting case of the 2HDM decoupling limit can
also be probed at the photon collider. The importance of the
sign of $m_{12}^2$ was studied in a more general context.
Clearly, positive  $m_{12}^2$ (in our notation) can lead to
large non-decoupling effects. Non-decoupling effects can also
appear due to higher order correction to the triple $h^0 h^0 h^0$
vertex. We have shown that non-decoupling effects will be more
easily seen in the low and in the high $m_\phi$ regions as in the
intermediate mass region the cross section is closer to the SM values.

In the fermiophobic limit this process is complementary to  the LEP
production process as it grows with $m_{12}$. Most importantly, it
can also probe a part of the parameter space that cannot be accessed
at hadron colliders. When $m_{12} \approx 0$ the cross section vanishes
at hadron colliders~\cite{Arhrib:2008pw}. On the contrary, we have
shown that in photon-photon collisions the cross section can reach
a few fbarn for  $m_{12} \approx 0$ and $\tan\beta\approx 5$. 
The region of
low $m_{12}$ in the fermiophobic scenario will most probably not be
excluded until we have access to a photon collider.

Regarding the CP-odd Higgs, we have shown that, for the energies
considered, only a light $A^0$ will be probed at a photon-photon
collider. For $m_{A^0} \ge 150$ $GeV$ the cross section is virtually
zero. We have also covered the mixed CP-even modes $h^0H^0$ and
$H^0H^0$ final state and checked that they can have comparable
cross sections to the $h^0h^0$ mode if $H^0$ is not too heavy. For the other modes, $h^0A^0$ and $H^0A^0$ we concluded that their cross section is at least one order of magnitude smaller
than the $h^0h^0$ and the $A^0A^0$ one.

As for the final states we have shown that in the two most
popular models, 2HDM-I and 2HDM-II, the $b \bar{b}$ 
final state is the preferred channel as its branching fraction is always at least the SM one. In this case and for not too heavy $h^0$ and $A^0$ a $4b$ final state can be searched for. We have showed the such a study for the SM can also be used for the 2HDM. This study can then be complemented with the final states $2b2\tau$  and $4\tau$. In other models and when other channels are kinematically available, a detailed study has to be performed taking into account all 2HDM parameters.


A last word: there is a well known complementarity between the LHC and the ILC. In this regard, if some of the 2HDM parameters and/or couplings are measured at the LHC and/or ILC
experiments such as Higgs masses and the magnitude of the vertices $h^0h^0h^0$ and $H^0h^0h^0$, this information could then be used at $\gamma\gamma$ experiments in order to
extract missing parameters and couplings like $h^0H^+H^-$ and  $H^0H^+H^-$.
\section{acknowledgments}
We would like to thank Fernando Cornet for valuable
discussions. C.C.H is supported by the National Science Council of
R.O.C under Grant \#s: NSC-97-2112-M-006-001-MY3 and R.B is
supported by National Cheng Kung University Grant No. HUA
97-03-02-063. R.S. is supported by the FP7 via a Marie Curie Intra
European Fellowship, contract number PIEF-GA-2008-221707.


\end{document}